# Visualizing and Quantifying Atomic Contributions to Raman Intensities governed by Spatially-Resolved Atomic Interferences

*Marc Bröckel[1,2], Johannes Gierschner[1,3], Alfred J. Meixner[1,2] and Kai Braun[1,2]**

[1]*Institute of Physical and Theoretical Chemistry, University of Tübingen, Auf der Morgenstelle 18, 72076 Tübingen, Germany*

[2]*Center for Light-Matter Interaction, Sensors&Analytics LISA+, University of Tübingen, Auf der Morgenstelle 15, 72076 Tübingen, Germany*

[3]*Madrid Institute for Advanced Studies, IMDEA Nanoscience, Madrid, Spain*

*Corresponding author: Dr. Kai Braun, kai.braun@uni-tuebingen.de

## Abstract

Raman spectroscopy is commonly reduced to molecular fingerprint sensing, neglecting Raman intensities. Atomic Raman intensity contributions trace Raman intensities back to their microscopic origin and act as local electronic structure descriptors; however, a both physical but intuitive framework is missing by now.

Therefore, in this work, we combine the two main lines of decomposing Raman intensities: atomic Raman tensors and atomic Raman Intensity Densities (RIDs). The former are used to define atomic Raman intensities which quantify both the magnitude and the phase of each atom's contribution to the global Raman signal, demonstrating that weak Raman bands may arise from destructive interference of individually strong atomic contributions. The latter build on redefined atomic Raman Polarizability Densities, RPDs, and are defined in analogy to our atomic Raman intensities. They show how the motion of an individual atom modulates the polarizability of the entire molecule. Atomic RIDs thus show local interferences of different atomic contributions. They integrate to the atomic Raman intensities bridging the two main lines. Additional (atomic) Charge Density Differences (CDDs) extend the RIDs to electronic structure effects. This new method provides indeed a physically motivated while chemically intuitive framework for interpreting Raman intensities. All required calculations are standard quantum chemical calculations and all necessary python scripts are published with instructions.

Finally, we apply the methodology to experimental surface-enhanced Raman spectra of substituted 2-mercaptobenzothiazole derivatives and discuss substituent-induced Raman intensity changes. We show that such changes do not necessarily stem from globally altered polarizabilities but from changes in the relative phase of atomic contributions. However, this method is not limited to SERS but widely applicable to all kinds of Raman, SERS or picocavity TERS experiments offering a basis for atomically resolved investigations of molecular processes such as adsorption, catalysis, and chemical reactions.

## Introduction

(Surface-enhanced) Raman spectroscopy combined with confocal microscopy has become a versatile and widely used method for probing molecular structures and dynamics with high spatial resolution as it has a high sensitivity for molecular geometries, chemical environments and various other external influences.[1–5] No external labelling or complex sample preparation is needed, which makes Raman spectroscopy widely applicable in diverse fields ranging from catalysis[6–10] and energy storage materials[11–14] to biomedical sensing[15–18] and environmental monitoring[19–21]. The strong increase in sensitivity in surface-enhanced Raman spectroscopy (SERS)[22] up to single-molecule detection[23–26] allows researchers to understand molecular processes in considerable detail.

Raman spectra are predominantly interpreted in terms of vibrational frequencies[1,2,5,22] whereas the intensities of Raman lines are often only qualitatively interpreted. Standard descriptions based on Raman tensors provide a global description of the system but do not resolve the atomic contributions, limiting mechanistic interpretations.[27,28] Recent advances in tip-enhanced Raman spectroscopy (TERS), particularly experiments exploiting plasmonic picocavities[29,30], have demonstrated that TERS can provide spatially resolved information beyond the conventional molecular average[31–33]. These experiments revealed that even individual functional groups or localized vibrational motions can be distinguished[26,26,34]. Such observations indicate that Raman intensities inherently contain information about the spatial origin of the polarizability modulation. However, a direct theoretical framework connecting these experimentally accessible local Raman responses to individual atomic contributions to the Raman signal is still lacking.

The interpretation of Raman intensities in terms of local contributions has evolved from empirical bond-based models to modern quantum-chemical decomposition approaches. The classical Bond Polarizability Model[35,36] (BPM) represents an early concept, attributing Raman intensity contributions to changes in the polarizability of individual chemical bonds. A spatially resolved description was later introduced by Chen and Liu[33] who define Raman Polarizability Densities (RPDs), which represent polarizability modulations occurring during normal mode oscillations as continuous functions in real space. Local integration yields their Locally Integrated RPD (LIRPD) approach with which they successfully resolve molecular structures using tip-enhanced Raman scattering (TERS) images[37]. Lee *et al.*[32] analogously consider polarizability modulations, however, they introduce an atomic decomposition using the Hirshfeld partitioning scheme[38] to decompose a molecule's global polarizability into atomic polarizabilities. With that, they obtain the change in atomic polarizability according to each atom's contribution to the ground state molecular electron density by numerical differentiation with respect to the normal mode. Such atomic RPDs are analogous real space polarizability modulation maps which the authors compare to their atomically confined TERS images. Their maps already include positive and negative atomic contributions to the polarizability modulation; however, their TERS images are proportional to the square of these atomic contributions effectively neglecting their relative phase information. They also use the Hirshfeld partitioning scheme for their atomic contributions which is an established partitioning scheme; however, it relies on an arbitrary definition of atoms in a molecule. Different approaches (e.g. the Quantum Theory of Atoms In Molecules, QTAIM[39], or the Voronoi partitioning scheme[40,41]) may define different atoms within the considered molecule yielding fundamentally different atomic contributions[42]. Chen and Jensen[43,44] and later Chaudhry *et al.*[45] build on the same Hirshfeld partitioning scheme of the molecular polarizability, defining their Raman Bond Model (RBM). Due to the high localization of the Hirshfeld partitioning of an atom's electron density to itself and its chemical bonds, Chen and Jensen decompose a molecule's Raman response into changes in the atomic induced charge densities and so-called Raman bonds (charge flow modulations between atoms). This yields not only a visual representation of contributing atoms and bonds but also lets them define and compute Raman spectra of e.g. adsorbate fragments. The RBM model takes also into account the phase information, i.e. whether the computed atomic or bond (or even fragment) contributions contribute constructively or destructively to the global Raman response. However, since they do not define exclusively atomic contributions, they cannot define purely atomic Raman spectra. To quantify a single atom's contribution to the global Raman intensity, many researchers followed a different approach in decomposing a normal mode's global Raman tensor into atomic Raman tensors[46,22]. This tensor decomposition is based on the decomposition of a normal mode into atomic motions. Consequently, atoms do not have to be arbitrarily defined but naturally arise from the derivatives of the molecular polarizability with atomic displacements. Specifically, with the development of atomically resolved TERS spectra, atomic Raman tensors are commonly discussed. Zhang *et al.*[47] model plasmonically enhanced Raman spectra by defining atomic Raman dipoles and weighting them with a strong, localized inhomogeneous electric field. Zhang *et al.*[34] further discuss those local Raman dipoles for interpreting local group vibrations observed in experimental TERS spectra. Schienbein[48] recently introduced a procedure which generates the polarizability autocorrelation function (from which the Raman intensity is obtained) from a molecular dynamics (MD) simulation. In this framework, the global polarizability change in time is partitioned into atomic contributions, allowing for an atomic decomposition of the polarizability autocorrelation function. One could thus define the atomic self-correlation as atomic Raman intensity, however, Schienbein considers large scale aqueous MD simulations and focuses on larger fragments (e.g. a sulfate anion).

In this work, we define atomic Raman intensities from the atomic Raman tensors, in analogy with how the global Raman intensity is calculated from the global Raman tensor. These atomic Raman intensities contain two types of information: their magnitude, representing the strength of the polarizability modulation caused by an atom, and their sign, reflecting the phase by which they couple to the global Raman intensity. This opens a new perspective on how Raman intensities arise. High Raman intensities result not only from high atomic Raman intensities but also from their constructive interference. Low Raman intensities, on the other hand, do not necessarily stem from atomic motions that hardly modulate the polarizability but can also occur when atoms contribute high polarizability modulations which, however, counteract one another. For a real-space representation of a normal mode's global Raman intensity, we define Raman intensity densities, RIDs, from RPDs analogous to those of Chen and Liu[33,37] with few corrections. Additionally, we define atomic RIDs not from an arbitrary and localized definition of atoms in a molecule but by defining atomic displacement vectors and generating single atomic displaced geometries. They show directly how the motion of one single atom modulates the polarizability of the entire molecule. Consequently, these local polarizability modulations not only entail information on their magnitudes but also on their local phase. Comparing different atomic RIDs reveals the atoms' contributions' local interference in real-space giving insights into local effects resulting in the global Raman intensity. These atomic RIDs integrate to our atomic Raman intensities, which makes them direct real-space representations thereof. With that, our method mathematically combines the previously separated main lines of decomposing Raman intensities: atomic RPDs/RIDs and atomic Raman tensors. Their combined discussion shows that high atomic Raman intensities are again expected not only for atomic motions leading to strong polarizability modulations but also constructive local polarizability modulations. Low atomic Raman intensities are observed for atomic motions leading to strong local polarizability modulations which, however, counteract one another. Furthermore, high atomic Raman intensities are not only observed for electron rich atoms but also atoms neighbouring electron rich regions as their motions can lead to strong electronic restructuring. Finally, we define normal mode (atomic) Charge Density Differences (CDDs) that depict the charge density oscillation occurring during a normal mode. The combination of CDDs and RIDs allows for a fundamental connection between a molecule's Raman intensity and its electronic structure. Finally, all atomic Raman intensity and atomic RID calculations are standard calculations and are performed with the standard quantum chemical program package Gaussian16. All python scripts generating (atomically) displaced molecular geometries and calculating atomic Raman intensities and (atomic) RIDs are published with extensive instructions.

Furthermore, we apply this method to interpret experimental SERS spectra of 2-mercaptobenzothiazole (MBT) and its derivatives 5-chloro-2-mercaptobenzothiazole (CMBT) and 6-ethoxy-2-mercaptobenzothiazole (EMBT) deposited on a thermally evaporated Ag-SERS substrate. These molecules share a common vibrational scaffold, while their charge densities are systematically altered through the substituents. Using this example, we can show the processes leading to high or low Raman intensities, which has not only to do with the strength of polarizability modulation but rather with constructive or destructive atomic contributions. Furthermore, due to the non-localized character of our (atomic) RIDs, we can also show that a molecule's normal mode not only modulates its own polarizability but also that of the SERS substrate. Consequently, a part of the measured Raman intensity stems from a polarizability modulation in the SERS substrate induced by the molecular vibration.

Importantly, our method does not rely on arbitrary definitions of atoms or empirical fitting parameters. Instead, it is derived directly from the Raman tensor formalism and the corresponding polarizability modulation during a normal mode cycle. Consequently, it is widely applicable and easily adjustable to all kinds of Raman, SERS or picocavity TERS experiments. Finally, this method paves the way for tracking electron density shifts with atomic resolution during processes such as adsorption, catalysis, chemical reactions, etc. as atomic Raman intensities directly translate global Raman intensities into atomic probes for the molecular electron density. Beyond immediate applications, the conceptual advance of atomically resolved Raman intensities motivates a re-examination of structure-property relations in molecular systems and provides experimentally accessible evidence thereof. This allows for a more detailed understanding yielding a high didactic value.

## Results and Discussion

The Raman effect is very sensitive to the electronic structure of a system as it describes scattering of photons at the system's electron cloud. Therefore, Raman spectroscopy can be used to track changes in the molecular charge distribution. Braun et al.[49] e.g. studied 2-mercaptobenzothiazole and benzotriazole employing tip-enhanced Raman spectroscopy (TERS) while applying an external electrical field via a bias between the tip and the substrate. They find no significant change in peak frequencies while the Raman intensities show a bias-dependent modulation. This is the first systematic study of how the Raman intensity directly reacts to an imposed shift in the molecular electron density. Here we extend this approach to intrinsic inductive and mesomeric effects. We compare the Raman spectra of three molecules of the 2-mercaptobenzothiazole (MBT) family as shown in Figure 1 a): the unsubstituted MBT, the 5-chloro-substituted CMBT and the 6-ethoxy-substituted EMBT. They all share the same MBT basis with similar Raman peak frequencies but different intensities. This allows us to systematically study the influence of the substituents on the molecules' electronic structure and on their Raman spectra

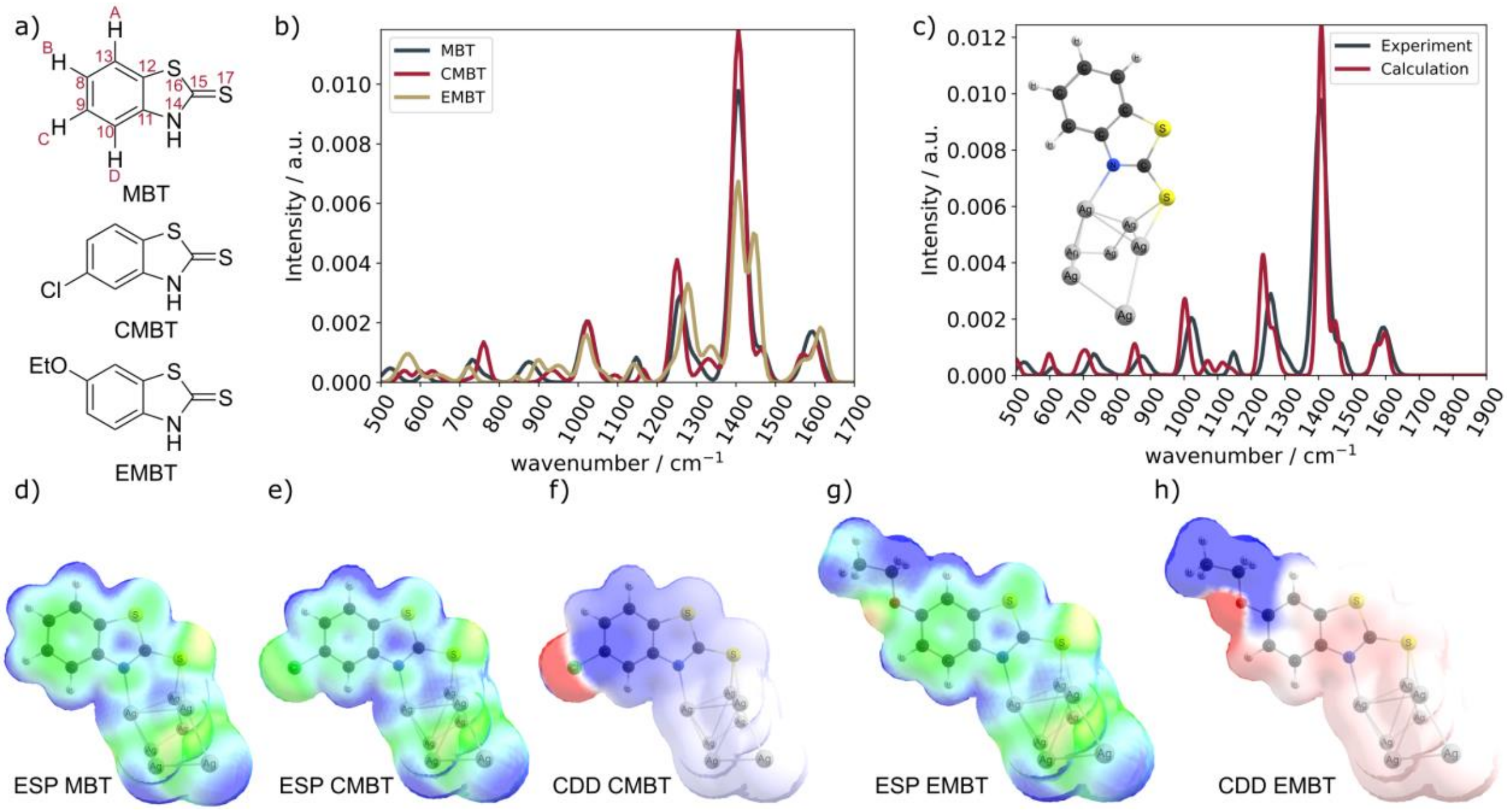

*Figure 1 a) shows the molecules MBT, CMBT and EMBT studied in this work. Their experimental SERS spectra on a rough Ag film are shown in b). c) Experimental and the calculated SERS spectrum of MBT. We find an excellent agreement both for the frequencies as well as for the Raman intensities which validates our computational method and allows for further theoretical considerations. The inset shows the adsorbate geometry used for the Raman calculation. d) Charge density distribution of the MBT-$Ag_7$-adsorbate model. Blue areas represent regions with a positive electrostatic potential (ESP) while red areas correspond to negative ESP regions. e) and g) show the ESP maps of CMBT and EMBT, respectively. f) and h) show the respective charge density differences (CDDs) of CMBT and EMBT with respect to MBT. With white being the indicator for a CDD of 0, we observe an electron deprived blue MBT basis for CMBT and a red electron rich MBT basis for EMBT. This highlights the dominant -I-effect for the Cl-substituent and the +M-effect for the EtO-substituent.*

The experimental SERS spectra of the three molecules shown in Figure 1 b) have been corrected for the instrumental background, the peaks were fitted with Gaussian functions, and the spectra were normalized to the spectral integral in the range from 500 – 3150 $cm^{-1}$ for comparability amongst the different molecules. We observe the expected modulation in Raman intensities, most prominent for the main feature at 1406 $cm^{-1}$ and the neighboring 1463 $cm^{-1}$ peak (a full band assignment can be found in chapter 1 of the SI). Thus, we will focus on these two peaks to discuss how different molecular electronic structures show in Raman intensities. We perform complementary DFT computations where we account for the Ag SERS substrate by means of the $Ag_7$-cluster shown in in the inlet in Figure 1 c). As shown in Figure S2, the 532 nm laser used for the SERS experiments is far off resonance for an electronic transition, so that electronic resonance effects may be neglected. The calculated SERS spectrum of the MBT-$Ag_7$-adsorbate shown in Figure 1 c) agrees well with the experimentally observed SERS spectrum, validating our computational approach for the following theoretical considerations (more details on the computations are given in chapter S4 in the SI; in chapter S5 we discuss the adsorption geometry of MBT on Ag in more detail.). Beginning with these adsorbate geometries, we can access their electronic properties. Figure 1 d) shows a map of the electrostatic potential (ESP) of the MBT-$Ag_7$-adsorbate. This plot was achieved by projecting the adsorbate's ESP onto its electron density with a fixed isodensity value of 0.01. Red and yellow areas show regions with an attractive ESP with respect to a positive probe charge while blue areas correspond to a repulsive ESP. The corresponding ESPs of CMBT and EMBT are shown in Figure 1 e) and g), respectively. We can further calculate charge density differences (CDDs) which show the differences in the molecules' electronic structures with respect to the parent molecule MBT. With white corresponding to a CDD of 0, blue areas now highlight areas with a more positive and red areas with a more negative ESP than MBT. The CDD of CMBT depicted in Figure 1 f) shows that the major influence of the Cl-substituent on the charge density originates from its electron-withdrawing inductive effect (-I) rooted in its high electronegativity. Electron density is being displaced from the MBT basis onto the Cl-substituent in no particular pattern; the inductive effect is solely limited by distance. The EtO-substituent in EMBT also shows an inductive effect on the bonding C-atom, however, it can be mainly characterized by its electron donating mesomeric effect (+M) which acts the strongest in ortho- and para-positions as shown in its CDD in Figure 1 h). We see that both substituents also influence the electronic structure of the $Ag_7$-cluster which makes it crucial for computational considerations. We note that, in comparing 5-CMBT and 6-EMBT, we mix electronic and structural properties. Therefore, we also calculated the ESPs of 5-ethoxy-2-mercaptobenzothiazole and 6-chloro-2-mercaptobenzothiazole (see Figure S6) and discuss their electronic properties in chapter S6.

As previously mentioned, we will focus on the two main features of the SERS spectra at 1406 $cm^{-1}$ and 1463 $cm^{-1}$ for our analysis as the substituent's effect on the Raman intensity is most pronounced there.

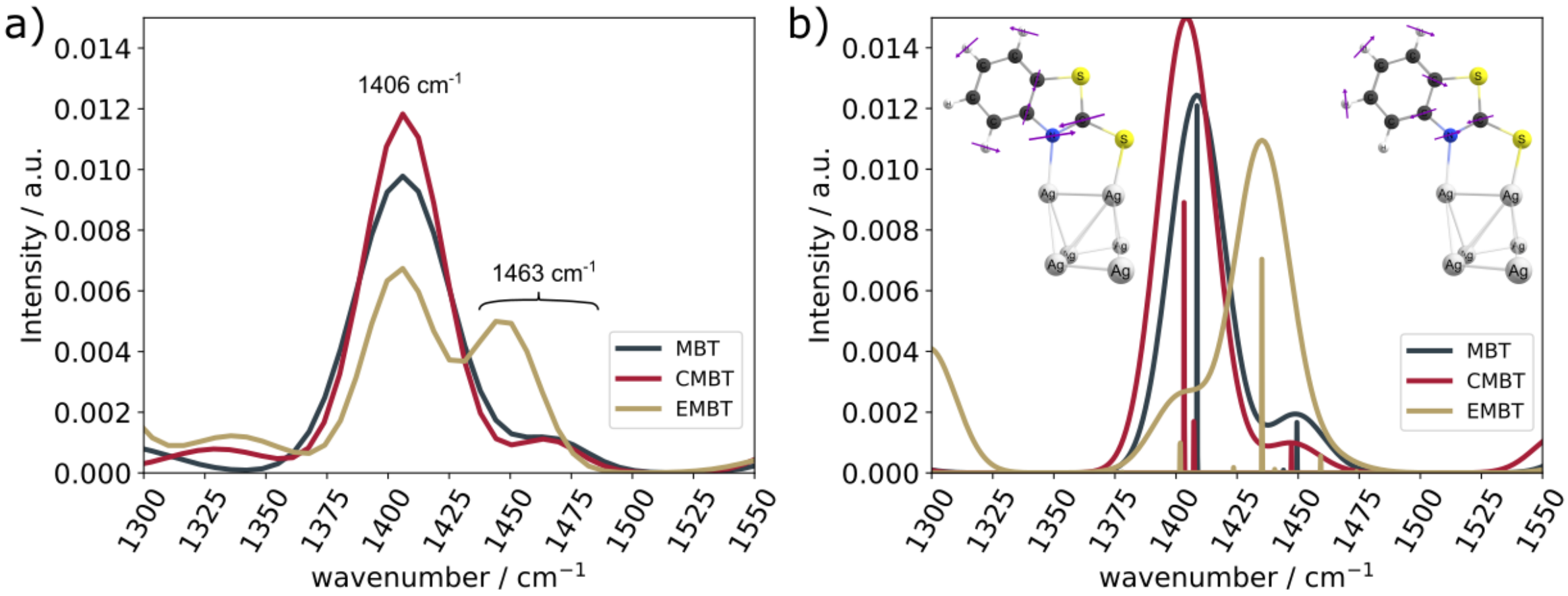

*Figure 2 a) zoom into the experimental SERS spectra of MBT, CMBT and EMBT to highlight the two studied peaks and to compare the experiments to the computed Raman spectra shown in b). Vertical lines show the normal modes which were folded with a Gaussian function with a FWHM of 20 $cm^{-1}$ to obtain the presented spectra. The insets show excerpts of the displacement vectors of the atoms involved in the normal modes mainly contributing to the*

Figure 2 a) shows a zoom into the experimental SERS spectra. For the 1406 $cm^{-1}$ peak, we observe an intensity sequence of CMBT > MBT > EMBT. The corresponding intensity sequence for the 1463 $cm^{-1}$ peak is EMBT > MBT > CMBT. Figure 2 b) shows our complementary quantum chemical calculations where the vertical lines show the normal modes and the enveloping spectra were obtained by convoluting the normal mode line spectrum with a Gaussian function with a FWHM of 20 $cm^{-1}$. The Raman intensity sequences are correctly reproduced for both peaks. The EMBT calculation correctly reproduces the relative peak intensity sequences for both peaks (as well as the order of magnitude) but incorrectly predicts a reversal of the main feature intensities. However, since we discuss the relative normal mode intensities in this work, we consider these calculations sufficient for our purposes.

We assign both experimental peaks to a single normal mode in order to systematically study the physical effects occurring during a normal mode cycle. While this can be unambiguously done for the 1463 $cm^{-1}$ peak as well as for MBT and EMBT for the 1406 $cm^{-1}$ peak, CMBT shows two normal modes close in energy that mix resulting in the high intensity of the 1406 $cm^{-1}$ peak. We compared both normal modes with their respective analogues in MBT and EMBT and find that the first of the two normal modes (with higher Raman intensity) describes the normal mode shown in the left inset in Figure 2 b). Therefore, we assign the 1406 $cm^{-1}$ CMBT peak primarily to the first of the two normal modes for comparability. That changes, however, the normal mode intensity sequence for the 1406 $cm^{-1}$ mode to MBT > CMBT > EMBT. Both studied normal modes exhibit a dominant C15-N14-stretching motion as shown by the normal modes' displacement vectors in the insets of Figure 2 b). The two modes differ in the coupling of the two bridging C-atoms C11 and C12. They stretch symmetrically and synchronously to the C-N-stretching the 1406 $cm^{-1}$ mode while they show more of a thiazole breathing character and move anti-symmetrically to the CN-stretching in the 1463 $cm^{-1}$ mode.

We now use a combined discussion of our atomic Raman intensities and (atomic) Raman polarizability densities to explain the different Raman responses of the three similar molecules. Atomic Raman intensities are obtained from the atomic Raman tensors (eq. (7)). Table 1 shows the calculated atomic Raman intensities for both modes where all atomic intensities have been scaled with a factor of $10^2$ for clarity (see Figure 1 a) for the numbering of the atoms). The contributions of the ethyl group in EMBT as well as the aromatic atoms A-D are not displayed as they are negligible in these two normal modes. The color map highlights significant positive contributions in blue and significant negative contributions in red.

*Table 1: Comparison of the atomic Raman intensities for the 1406 $cm^{-1}$ and 1463 $cm^{-1}$ modes in MBT, CMBT and EMBT. High positive values are highlighted in blue while high negative values are highlighted in red. The values have been scaled with a factor of $10^2$ for clarity.*

| | 1406 $cm^{-1}$ | | | 1463 $cm^{-1}$ | | |
|---|---|---|---|---|---|---|
| | MBT | CMBT | EMBT | MBT | CMBT | EMBT |
| I | 1.24 | 1.16 | 0.25 | 0.18 | 0.10 | 1.05 |
| C8 | 0.00 | 0.00 | 0.02 | -0.01 | 0.01 | -0.08 |
| C9 | 0.00 | -0.01 | 0.01 | 0.00 | 0.01 | -0.02 |
| C10 | 0.02 | 0.00 | 0.01 | 0.00 | 0.00 | 0.01 |
| C11 | 0.19 | 0.08 | 0.01 | 0.00 | -0.05 | 0.19 |
| C12 | 0.11 | 0.07 | -0.03 | 0.02 | -0.02 | 0.19 |
| C13 | -0.03 | 0.00 | -0.04 | 0.02 | 0.01 | 0.02 |
| N14 | 0.40 | 0.39 | 0.10 | 0.08 | 0.06 | 0.35 |
| C15 | 0.57 | 0.62 | 0.16 | 0.08 | 0.08 | 0.39 |
| S16 | 0.00 | 0.00 | 0.00 | 0.00 | 0.00 | 0.00 |
| S17 | 0.01 | 0.01 | 0.00 | 0.00 | 0.00 | 0.00 |

Atomic Raman intensities contain two sets of information: the magnitude, indicating the contribution of each atom to the global Raman intensity, and their sign, which represents the phase of each atomic contribution with respect to the global Raman intensity. High values can be expected for electron rich atoms (e.g. N14 in the MBT family) as their motion is likely to lead to a large redistribution of the electron density which then goes into the change in polarizability. However, atoms neighboring electron rich regions (e.g. C15) can also strongly influence the Raman response as their motions can also lead to strong electronic restructuring. Atomic Raman intensities can be negative per definition which, however, does not correlate to measurable negative intensities. Atomic contributions, which couple in-phase to a common change in polarizability, show the same sign in their atomic Raman intensities. Atomic contributions that counteract one another yield a low net polarizability change which is reflected in opposite signs in their atomic Raman intensities. However, negative atomic Raman intensities are always comparably small as the sum over all atomic intensities must always yield the positive square of the reduced trace of the global Raman tensor.

Atomic Raman intensities are already an interesting tool; however, they lack a chemically intuitive interpretation. Raman Intensity Densities (RIDs) are real-space molecular maps showing the polarizability modulation occurring during a normal mode (eq. (19)). Additionally, we define atomic RIDs from atomic displacement vectors whose sums yield their respective global values. They do not rely on any arbitrary definition of atoms in a molecule and show how the polarizability of the entire molecule is modulated by the single atom's motion. They integrate to the atomic Raman intensities (except for numerical noise) and are consequently a real-space representation of the atomic Raman intensities which allows for a detailed interpretation of local polarizability effects. Additionally, we define (atomic) normal mode CDDs as difference between the ESPs of the two normal mode extremal structures (shown in Figures S10 and S11). While RIDs are always identical for both normal mode half cycles, the CDDs continuously reverse their signs. We define all CDDs discussed in this work such that they always show the charge flow during C15-N14 bond elongation. The comparison between the (atomic) normal mode RIDs and CCDs allows for a fundamental discussion of the origin of the normal mode's Raman intensity. Figure 3 shows the decomposition of the MBT Raman spectrum into atomic contributions and the atomic RIDs of the 1406 $cm^{-1}$ mode.

### The 1406 $cm^{-1}$ peak

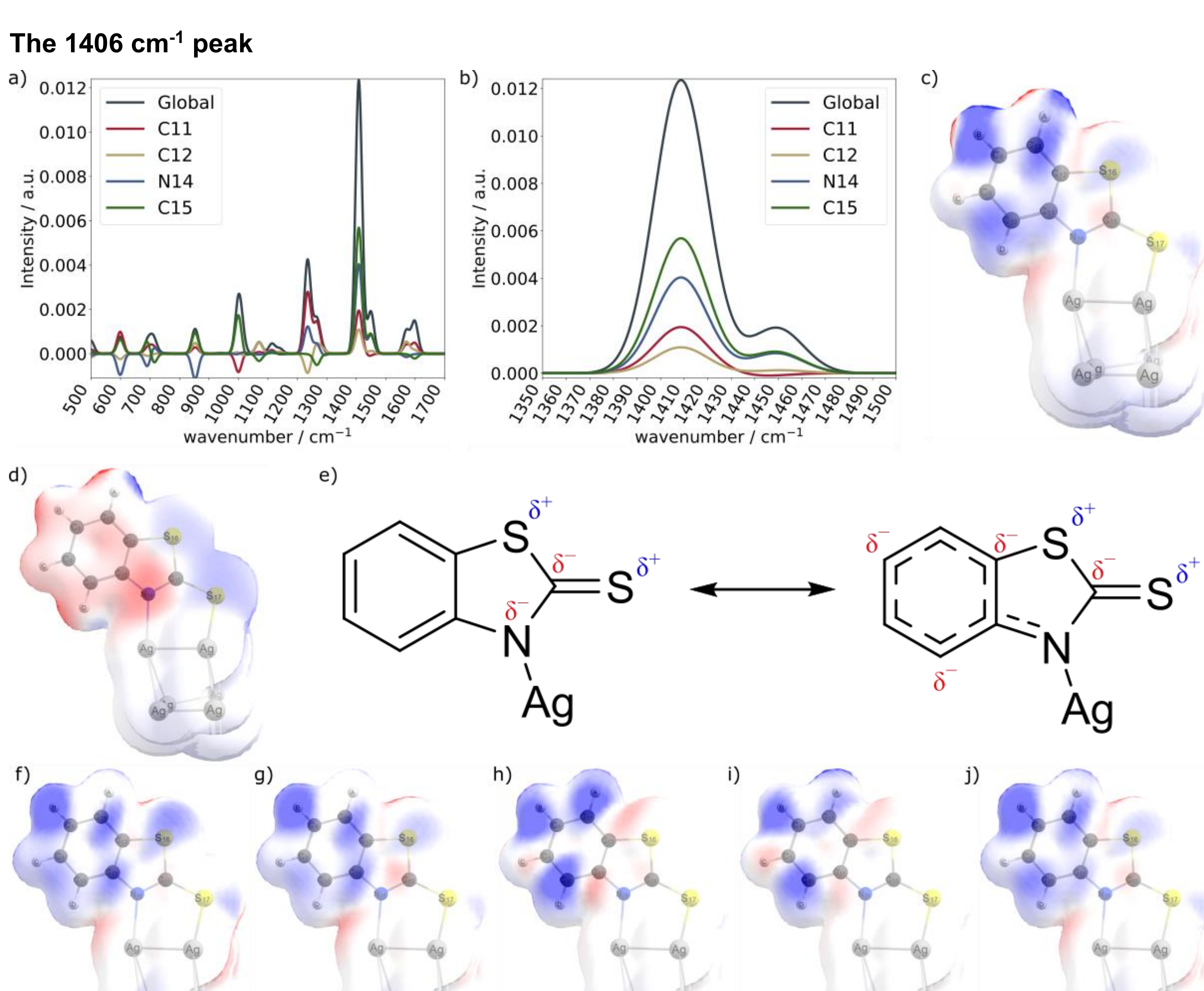

*Figure 3 a) shows the global computed MBT Raman spectrum in black and the atomic spectra of C11, C12, N14 and C15 which are the main contributors for most peaks. Atomic Raman intensities can be negative due to their definition, however, negative contributions are small compared to positive contributions and always result in a global positive Raman intensity. b) shows a zoom into the two considered peaks at 1406 $cm^{-1}$ and 1463 $cm^{-1}$. c) shows the global Raman Intensity Density (RID) of the 1406 $cm^{-1}$ normal mode. The color scale is set to $\pm 2.5 \cdot 10^{-5}$ a.u. Å$^{-3}$. d) shows the Charge Density Difference (CDD) that occurs during the normal mode oscillation (color scale $\pm 0.05$ a.u.). e) gives a representation in resonance Lewis structures thereof. The C15-N14 bond elongation induces a partial negative charge on N14 that is delocalized in the benzene moiety via the mesomeric effect. f) – i) show the atomic RIDs of C15, N14, C11 and C12. The color scales are set to $\pm 10^{-5}$ a.u. Å$^{-3}$ for C15 and N14 and $\pm 5 \cdot 10^{-6}$ a.u. Å$^{-3}$ for C11 and C12. j) shows the sum of the atomic RIDs shown in f) – i) and the scale is set to $\pm 2.5 \cdot 10^{-5}$ a.u. Å$^{-3}$.*

Table 1 shows that C11, C12, N14 and C15 are the main contributors to the 1406 $cm^{-1}$ and the 1463 $cm^{-1}$ modes. Figure 3 a) and b) thus show the decomposition of the global Raman spectrum of MBT into the atomic contributions of C11, C12, N14 and C15. The spectra were obtained by convoluting the atomic Raman intensity line spectra with a Gaussian function with a FWHM of 20 $cm^{-1}$. Figure 3 c) shows the global RID for the 1406 $cm^{-1}$ mode of the MBT-$Ag_7$-adsorbate. Blue areas (in Figure 3 c – h) represent a positive RID and thus an increase of the polarizability during the normal mode while red areas show a negative RID and hence a decrease. Since RIDs integrate to the global Raman intensity, modes with predominant positive (blue) RID features show high Raman intensities while in those with equal positive and negative contributions, the RID areas cancel yielding low intensity modes. We see in Figure 3 c) that the main features of the 1406 $cm^{-1}$ mode's Raman intensity are located on the C8-B bond, the C10-C11 bond, C13 and S16 with only little destructive (red) RID areas around H atoms A and B as well as C15 and the N14-Ag bond. Furthermore, the RID is not only localized to the MBT molecule but extends into the $Ag_7$-cluster. The electronic structure of the SERS substrate near the molecule is thus influenced by a normal mode of the adsorbed molecule and contributes to the molecule's Raman intensity. This is an enhancement mechanism not commonly discussed besides the electromagnetic and the chemical enhancement. Persson *et al.*[51] first derived this effect in 2006 for metallic nanoparticles and Chen and Jensen[43] used their Raman bond model to demonstrate that the Raman intensity of an adsorbate's normal modes is strongly rooted in the Ag cluster used in their calculations. Interestingly, the main fraction of Raman intensity arises in the benzene moiety even though the dominating atomic motions (especially the C15-N14 bond stretching) occur in the thiazole moiety. This can be rationalized with the normal mode CDD shown in Figure 3 d). It shows that C15-N14 bond elongation induces a partial negative charge on N14 and the benzene moiety, as well as a positive charge on the two S atoms. This, in turn, can be understood by examining the resonance Lewis structures shown in Figure 3 e) (the corresponding natural population analysis is shown in SI chapter S11). The C15-N14 bond is polarized toward N14 due to its higher electronegativity. NPA shows that bond elongation further reduces the influence of C15 on the bond electrons which are then, in turn, further polarized onto N14 inducing a partial negative charge. The resulting electron deficiency on C15 is overcompensated by the two neighboring S atoms and they both acquire a partial positive charge (while they induce a partial negative charge on C15). The electronic restructuring of the S16-C15-S17 unit is associated with a large shift in electron density; however, this shift is highly localized. Consequently, the molecule's total polarizability changes only slightly, so this isolated CDD has no significant effect on the RID. The negative charge on N14, on the other hand, is strongly delocalized into the benzene's $\pi$-system via mesomeric conjugation. It is strongest in ortho- and para-positions which are also the locations with the highest RID features. This dominant contribution to the RID can therefore be attributed to both the magnitude of induced charge redistribution as well as the coupling to the highly polarizable, delocalized $\pi$-system. This interpretation of the CDD can be further explained employing Weinhold's NBO method[50] which converts the molecular orbitals into natural bond orbitals (NBOs) yielding a Lewis-like description of the electronic structure (see SI chapter S12). Second-order perturbation theory can then be used to calculate the energies of NBO donor-acceptor interactions, E(2). Large changes during normal mode oscillations, thus large $\Delta$E(2) values, indicate that a mesomeric resonance changes significantly during a vibration, which is typically accompanied by a substantial change in the electronic structure. Considering $\Delta$E(2) for the N14 $\rightarrow \pi^*$, S16 $\rightarrow \pi^*$ and later X $\rightarrow \pi^*$ (with X being Cl in CMBT and O in EMBT) interactions consequently allows for a mechanistic interpretation showing which orbitals induce the observed CDDs. In the case of MBT, the N14 $\rightarrow \pi^*$ interaction increases upon C15-N14 bond elongation, consistent with an increase in delocalization of the induced negative charge on N14 into the benzene moiety.

The atomic RIDs shown in Figure 3 f) – i) further decompose the global Raman effect in relating the motion of a single atom to its contribution to the global polarizability modulation during the normal mode. C15 has the largest atomic Raman intensity and its RID contributes the main features to the global RID. N14 has the second largest positive atomic Raman intensity where its positive sign indicates constructive interference to the contribution of C15. This can be illustrated by their atomic CDDs and $\Delta$E(2) values: their CDDs show that their individual motions induce the partial negative charge on N14 while also increasing the N14 $\rightarrow \pi^*$ interaction leading to its delocalization into the benzene's $\pi$-system. C11 and C12 show further positive atomic Raman intensities showing constructive coupling of their induced polarizability modulations to the global polarizability modulation. The corresponding atomic CDDs and $\Delta$E(2) values show that their motions modulate the mesomeric resonances of the N14 $\rightarrow \pi^*$ and S16 $\rightarrow \pi^*$ interactions. During C15-N14 bond elongation, C11 moves closer to N14 increasing the overlap of its lone pair and the benzene's $\pi$-system, resulting in a charge flow from N14 into the benzene moiety. Analogously, C12 moves closer to S16 upon C15-N14 bond elongation delocalizing more of the S16's lone pair into the benzene moiety. Consequently, both atomic motions further modulate the electron density of the benzene's $\pi$-system, resulting in a larger modulation of its polarizability and thus contribute significantly to the global Raman intensity. However, in contrast to the C15 and N14 motions, which delocalize an induced partial charge, the C11 and C12 motions modulate the delocalization of the lone pairs of N14 and S16 inducing partial positive charges on them. This leads to a local decrease in polarizability modulation, inducing negative, destructive RID features. These areas locally interfere destructively with the C15 and N14 RIDs, lowering the global polarizability modulation which explains their lower atomic Raman intensities.

Combining our considerations of atomic Raman intensities, atomic RIDs and atomic CDDs unravels the fundamental processes governing the experimentally observed Raman intensity of MBT. We now apply this formalism to explain the experimentally observed decreased Raman intensities in CMBT and EMBT. Their atomic Raman spectra as well as their global and atomic RIDs are shown in Figure 4.

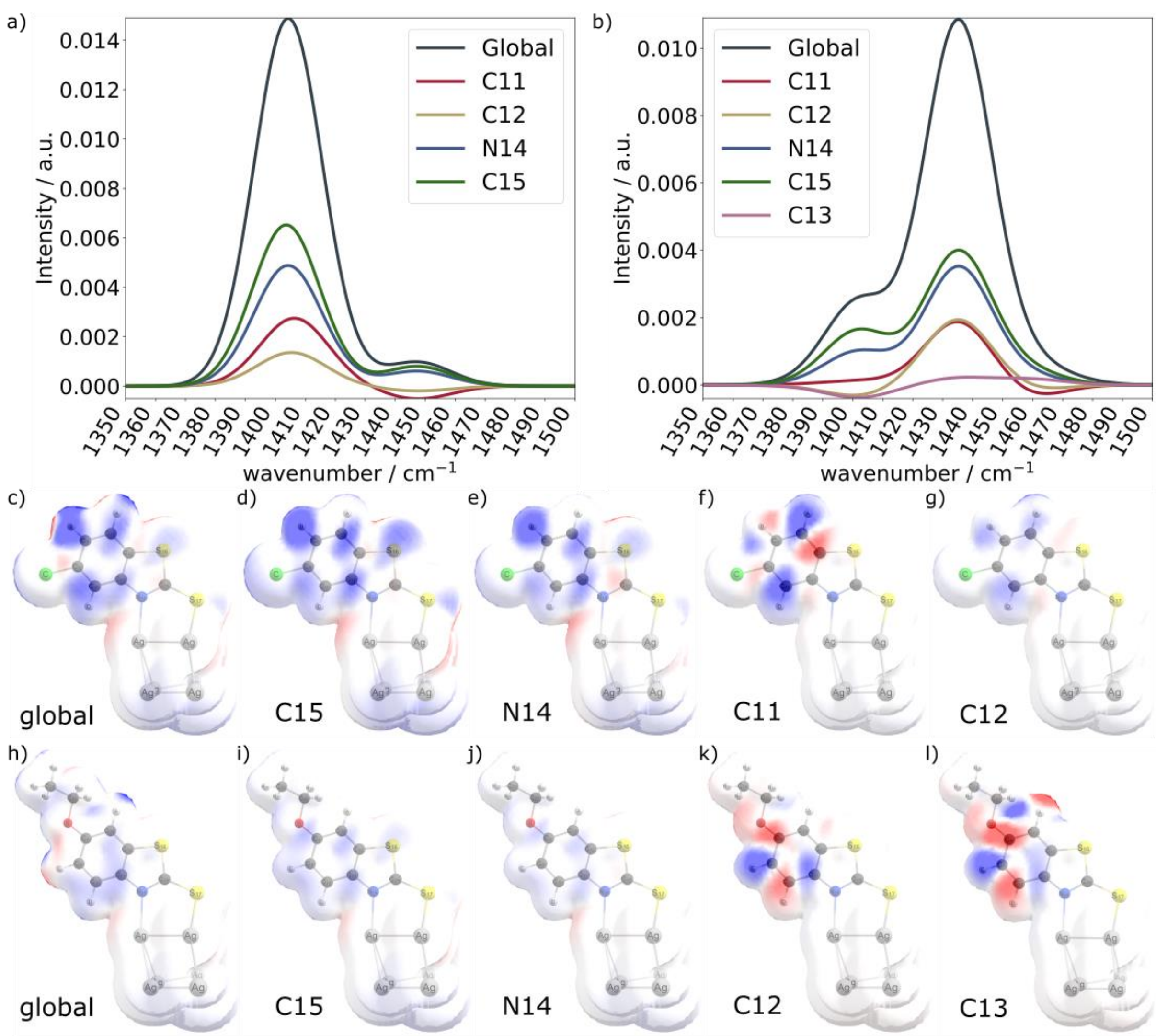

*Figure 4: a) shows the atomic Raman spectra of CMBT and b) that of EMBT. c) shows the global normal mode RID of the 1406 cm$^{-1}$ CMBT mode with the atomic contributions d) – g). The color map of the global RID is limited to $\pm 2.5 \cdot 10^{-5}$ a.u. Å$^{-3}$, the C11 and C12 contributions to $\pm 5 \cdot 10^{-6}$ a.u. Å$^{-3}$ and the N14 and C15 contributions to $\pm 5 \cdot 10^{-6}$ a.u. Å$^{-3}$. Analogously, h) shows the global RID for the EMBT 1406 cm$^{-1}$ mode with its atomic contributions i) – l). The color scale is $\pm 10^{-5}$ a.u. Å$^{-3}$ for the global RID, $\pm 5 \cdot 10^{-6}$ a.u. Å$^{-3}$ for the C15 and N14 and $\pm 2.5 \cdot 10^{-6}$ a.u. Å$^{-3}$ for the C12 and C13 contributions.*

The global RID of CMBT shows similar characteristics and is comparable in magnitude to that of MBT, which is consistent with the similar global CDD. It similarly describes an induced negative charge on N14 (upon C15-N14 elongation), which is delocalized into the benzene moiety via mesomeric resonances. Additionally, a negative charge is induced on the Cl-substituent which particularly shows in the atomic CDDs of C15 and N14 upon C15-N14 elongation. This increases the polarizability on the substituent as shown in the positive features in their RIDs resulting in C15 having a higher atomic Raman intensity in CMBT than in MBT (while that of N14 is nearly equal). The induced negative charge on the Cl-substituent is consistent with its electron-withdrawing -I-effect rooted in its high electronegativity. At the same time, the reduced Cl → π* interaction indicates a decreased mesomeric conjugation of the substituent with the benzene's π-system. Both effects contribute to the observed induced negative charge and the associated polarizability modulation. The C11 induced CDD shifts in comparison to MBT from the modulation of the N14 lone pair to a CDD completely localized within the benzene ring. This is likely caused by the influence of the Cl-substituent's mass on the normal mode. While C11 moved towards N14 in MBT, it moves parallel to N14 in CMBT. Consequently, the electron density of the benzene's π-system is modulated due to the C11 motion, however, it is only shifted between the upper and the lower half of the benzene ring. No additional electron density of the thiazole hetero atoms is being coupled. The C12 motion retains its movement in CMBT and thus modulates the overlap of the lone pair of S16 and the benzene ring. Interestingly, it shows a similar S16 $\rightarrow \pi^*$ interaction strength modulation (with no Cl participation; $\Delta$E(2) Cl $\rightarrow \pi^*$ is nearly zero) while it has significantly weaker CDD and RID. The reduced CDD and RID indicate that the difference arises from a different spatial electronic response of the $\pi$-system instead of different mesomeric modulation strengths. In summary, the Cl-substituent's -I-effect as well as its reduced mesomeric electron-donation increases the charge oscillation of the induced negative charge on N14 and the benzene moiety increasing the corresponding polarizability modulation (C15 and N14 contributions), its electronic influence on the MBT's electron density reduces the charge density oscillation between S16 and the benzene moiety (C12 contribution) and its mass alters the normal mode displacement of C11, thereby suppressing a charge density oscillation between N14 and the benzene moiety (C11 contribution), decreasing the global polarizability modulation. The atomic Raman intensities quantify the influences of these effects onto the global Raman intensity and show that the decreases in polarizability modulations overcompensate the increasing effects resulting in the experimentally measured lower Raman intensity of the 1406 $cm^{-1}$ peak of CMBT in comparison to the MBT.

In EMBT, the C15-N14 bond elongation also induces a negative charge on N14 that is delocalized into the benzene ring; however, this CDD and the associated effect on the polarizability are strongly decreased in comparison to MBT and CMBT. This can be rationalized by the EtO-substituent's strong electron-donating +M-effect. During C15-N14 bond elongation, the O $\rightarrow \pi^*$ interaction is increased, coinciding with a stronger mesomeric delocalization of its lone pair in the $\pi$-system. This additional electron donation counteracts the delocalization of the induced charge from N14 into the benzene moiety, thereby significantly weakening the associated charge shift. This accounts for the lower global CDD as well as C15 and N14 CDDs which correlates with the lower RIDs and thus Raman intensity. It is also evident in a change in the normal mode displacement pattern. While the C15-N14 bond is being modulated by $0.32$ Å in MBT and $0.33$ Å in CMBT, it is modulated by only $0.20$ Å in EMBT. Consequently, the motions of the other atoms are also influenced. C11 moves parallel to N14 inducing a CDD that is confined within the benzene moiety. Its RID contains nearly equal positive and negative features so that integration yields a contribution of nearly zero. C12 still modulates the mesomeric effect of S16, however, here again, the strong mesomeric effect of the EtO-substituent suppresses additional charge flows into the benzene's $\pi$-system and strongly localizes it onto C13. The corresponding C12 RID now shows more negative (red) than positive (blue) areas so that the C12 motion leads to a dominating global decrease in polarizability modulation. Its atomic Raman intensity is negative, showing its destructive contribution to the global Raman intensity. The altered normal mode displacement pattern gives rise to a new non-negligible atomic contribution that was not relevant before: The C13 atom, which moves towards the EtO-substituent. Its motion yields a similar RID to that of C12 which is also negative, lowering the global Raman intensity.

In summary, the 1406 $cm^{-1}$ mode involves three major charge density redistribution pathways from the thiazole heteroatoms into the benzene's $\pi$-system mediated by different atomic motions. The C15-N14 bond elongation (thus the C15 and N14 motions) induces a negative charge on N14 which can then be delocalized into the benzene ring. C11 modulates the intrinsic M-effect of the N14 lone pair, while C12 modulates that of the S16 lone pairs. The electron withdrawing -I-effect of the Cl-substituent promotes the charge density shift of the induced negative charge of N14 into the benzene moiety while its influence on the MBT's electronic structure alters the C11 and C12 induced CDDs and thus RIDs. In contrast, the strong electron donating +M-effect of the EtO-substituent in EMBT counteracts additional charge flows into the benzene's $\pi$-system in this mode, thereby reducing its polarizability modulation and thus Raman intensity. This significant reduction in Raman intensity in comparison with MBT or CMBT is directly reflected in the experimental spectra.

Chen and Jensen[43] similarly employ their Raman Bond Model (RBM) to discuss the influence of substituents to the global Raman intensity. Their RBM uses a Hirshfeld partitioning of the molecular polarizability and they calculate atomic and bond contributions to the Raman intensity using FEM differentiation along a normal mode. They also show that a substituted molecule's decreased Raman intensity can be caused by a change in contribution phases, however, their interpretation is more focused on the conjugation of their Raman bonds rather than the local interference of polarizability modulations caused by the atoms.

**The 1463 cm$^{-1}$ peak**

In analogy to the previous discussion of the 1406 cm$^{-1}$ mode, the experimentally observed Raman intensities of the 1463 cm$^{-1}$ peak can be rationalized with atomic RIDs and CCDs (Figures S9 and S11). The calculated spectrum in Figure 2 b) correctly reproduces the experimentally observed Raman intensity sequence of EMBT > MBT > CMBT. While the 1406 cm$^{-1}$ mode is predominantly governed by the C15-N14-stretching, the 1463 cm$^{-1}$ mode also features a C15-N14-stretching motion, however, it cannot be as easily reduced to that. The modulation of the C15-N14 bond length is decreased in MBT ($\Delta d = 0.15$ Å) and CMBT ($\Delta d = 0.18$ Å) but increased in EMBT ($\Delta d = 0.28$ Å). The benzene moiety now shows a stronger ring breathing character, and its bond lengths are more strongly modulated than in the 1406 cm$^{-1}$ mode. Thus, while the 1406 cm$^{-1}$ mode was determined by charge density flowing from the thiazole to the benzene moiety, the 1463 cm$^{-1}$ mode is more determined by a charge density oscillation within the benzene moiety. Furthermore, the C11 and C12 displacement directions remain similar in the 1463 cm$^{-1}$ mode, their phases change. While the C11-C12 bond stretched simultaneously with the C15-N14 bond in the 1406 cm$^{-1}$ mode, it compresses during the C15-N14-stretching (and vice versa).

C15 and N14 remain the main contributors to the 1463 cm$^{-1}$ normal mode with similar characteristics to the 1406 cm$^{-1}$ mode in all three molecules. The C15-N14 bond elongation induces a negative charge on N14 that is being delocalized into the benzene's $\pi$-system. This modulates the polarizability of the benzene's $\pi$-system providing the basis for this mode's Raman intensity. Due to the lower C15-N14 bond length modulations in MBT and CMBT, the associated polarizability modulation is decreased in their respective 1463 cm$^{-1}$ normal modes. The respective atomic Raman intensities of C15 and N14 are significantly decreased. In contrast, the modulation of the EtO-substituent's mesomeric coupling reverses compared with the 1406 cm$^{-1}$ mode. The O $\rightarrow \pi^*$ interaction is significantly decreased allowing for a significantly increased N14 $\rightarrow \pi^*$ interaction.

As previously mentioned, C11 and C12 change the direction of their motion in the 1463 cm$^{-1}$ mode. In MBT, even though C11 reverses its direction, it reduces its distance to N14 during the C15-N14 bond elongation causing a charge density shift from the thiazole to the benzene moiety. However, while the mesomeric effect of N14 induced a partial negative charge on C13 and a partial positive charge near C10 in the 1406 cm$^{-1}$ mode, these CDD features reverse in the 1463 cm$^{-1}$ mode. This CDD reversal is accompanied by a corresponding RID reversal on C10 and C13 so that the C11 RID for the 1463 cm$^{-1}$ mode shows no longer mainly positive features but equal positive and negative features which cancel upon spatial integration yielding an atomic Raman intensity of zero. The reversal of the motion direction of C12 directly reverses its entire CDD, now shifting electron density from C9, C11 and C13 (ortho- and para-positions of S16 in the benzene moiety) onto S16. The corresponding RID map shows, like that of C11, nearly equal positive and negative features also integrating to an atomic Raman intensity of nearly zero. In CMBT, the reversal of the C11 and C12 directions directly translate to reversed CDDs and reversed RIDs. Consequently, their atomic Raman intensities become negative, reducing the global Raman intensity. Finally, C11 strongly moves towards N14 in the 1463 cm$^{-1}$ EMBT mode which induces the previously discussed charge flow from N14 into the benzene moiety. It is less hindered in comparison to its 1406 cm$^{-1}$ mode due to the simultaneous reduction of the EtO-substituent's mesomeric electron-donation yielding a high polarizability modulation. Lastly, C12 induces a charge density flow from C13 into the lower half of the benzene moiety as well as S16 strongly modulating the polarizability of the benzene's $\pi$-system. The C11 and C12 RIDs now show dominant constructive areas so that both atoms become important contributors to the global Raman signal.

## Conclusions

Our work establishes a new direct and quantitative link between Raman intensities and the underlying electron dynamics in molecules. By combining (surface-enhanced) Raman spectroscopy with state-of-the-art quantum chemical simulations, we demonstrate that Raman Intensity Densities (RIDs) are an effective tool to visualize the physical processes governing the Raman response of a molecule. Furthermore, we establish a link between a molecule's RID and its normal mode charge density (CDD) so that the Raman response of a system can be broken down into electronic effects with atomic resolution. Notably, because these (atomic) RIDs are inherently non-local, they also reveal that part of the Raman intensity of an adsorbed molecule originates from a polarizability modulation induced in the SERS substrate itself – a contribution that is invisible to purely molecule-centered descriptions. These findings challenge the conventional view of Raman intensities as phenomenological observables derived from molecular polarizability tensors and instead promotes them as quantitative indicators of electron structure changes.

The introduction of atomic Raman intensities provides a powerful tool for decomposing global Raman spectra into chemically intuitive atomic contributions. Crucially, these atomic Raman intensities carry not only a magnitude but also a phase, so that a molecule's global Raman intensity emerges from local constructive or destructive interference of its atomic contributions rather than from their magnitudes alone. We present a new method for defining atomic RIDs from atomic displacement vectors rather than from an arbitrary definition of atoms in a molecule. They are inherently non-local and show their polarizability modulation in the entire molecule. Consequently, atomic RIDs, which integrate to the atomic Raman intensities, provide a direct real-space visualization of local atomic interferences. Raman intensities are governed not only by the individual strengths of atomic polarizability modulations but their interference. High local atomic contributions can result in low Raman intensities when they interfere destructively. This atomistic approach thus offers an intuitive interpretation of chemical effects, e.g. substituent effects or conjugation – as demonstrated for MBT and its derivatives CMBT and EMBT – in terms of their direct influence on local charge density and interatomic interference.

Beyond its conceptual significance, this methodology enables the rational prediction and tuning of the Raman response through electronic structure design, providing new pathways for optimizing materials for sensing, catalysis or photonic applications. Because atomic Raman intensities and atomic RIDs directly translate global Raman intensities into atomic-resolution probes of the molecular electronic structure, this formalism further paves the way for tracking electron density shifts during processes such as adsorption, catalysis, or chemical reactions. Consequently, this formalism goes hand in hand with time- or spatially resolved spectroscopies where electronic structure tracking could provide unprecedented insights into molecular mechanisms and electronic structure dynamics. Such an approach could enable design strategies where desired Raman responses guide molecular design with tailored electronic properties.

In summary, by decomposing Raman spectra into atomic contributions without relying on an arbitrary definition of atoms and connecting them to a molecule's electronic structure, this work bridges the gap between Raman spectroscopy and electronic structure theory. It expands Raman spectroscopy from a technique to qualitatively match reference spectra into a chemically intuitive, quantitative probe of local electronic behaviour. This framework not only introduces a new perspective to the Raman process, centered on the constructive and destructive interference of atomic contributions, but also establishes new possibilities for exploiting Raman scattering as a window into the electronic structure of matter.

## Methods

### SERS Experiments

SERS substrates were produced by evaporating (Pfeiffer PLS570, $10^{-6}$ mbar, 0.1 nm/s) 7 nm thick Ag-island films on plasma-cleaned (Diener Femto, 60 min, $O_2$-plasma, p=0.3 mbar) glass cover slips. 2-mercptobenzothiazole (MBT), 6-Ethoxy-2-mercaptobenzothiazole (EMBT) and 5-Chloro-2-mercaptobenzothiazole (CMBT) were purchased from SigmaAldrich and used without further purification. The SERS samples were each placed for 30 min in $10^{-3}$ M solutions of the molecules and subsequently intensively rinsed with methanol. Comparison with bulk Raman spectra (showing dimerization) ensured the formation of self-assembled monolayers (SAMs).

SERS spectra were recorded with a home-built inverted confocal microscope equipped with a Zeiss $\alpha$ Plan-APOCHROMAT 1.46 NA oil objective, a 532 nm continuous-wave excitation laser (coherent compass 215M) and a Princeton Instruments Acton SP2500 spectrometer (grating density 150 g/mm) followed by a ProEM 512 EMCCD detector. The laser power ranged between 0.2 mW and 2 mW and the acquisition time was adjusted accordingly to between 1 s and 60 s to prevent molecular pyrolysis.

1500 spectra were collected for MBT, 3000 for EMBT and 4500 for CMBT. They were averaged and instrumental background corrected. The observed peaks were fitted with Gaussian functions. Since the measurements were conducted over a longer period, the setup slightly misaligned, which shifts the Raman bands by a max. of 10 $cm^{-1}$. This would artificially broaden the peaks during averaging so that we set the position of the major peak to 1406 $cm^{-1}$ and shifted all spectra accordingly. The averaged spectra were normalized to the total integral of the corresponding spectrum in the region 500 – 3150 $cm^{-1}$ to enable comparability between different molecules. Data analysis was performed using custom Python scripts.

### DFT Computations

All quantum chemical calculations presented in this work were performed with the Gaussian16 program package[52]. Following the cluster ansatz first established by Zhao et al.[53] and Wu et al.[54] to incorporate the chemical effect of the nearest silver atoms to the molecules, we found a $Ag_7$ cluster that correctly reproduced the experimental spectra (multiple MBT-$Ag_n$-adsorbates were considered and this MBT-$Ag_7$-adsorbate showed the best match between calculated and experimental SERS spectrum; see SI chapter S4). Adsorbate structures were obtained using the long-range-corrected CAM-B3LYP[55] hybrid functional corrected with Grimme's D3[56] dispersion correction and Becke-Johnson damping[57]. We used the Stuttgart-Dresden (SDD) MWB28[58,59] effective core potential for the Ag atoms and the cc-pVTZ basis set[59–62] for the other elements. Frequency calculations confirmed minima in the adsorbates' PESs and were used to calculate Raman spectra. The harmonic frequencies are scaled by a scaling factor of 0.954[63] and broadened with a Gaussian function with a FWHM of 20 $cm^{-1}$ for all these adsorbates. We normalized all calculated spectra to the spectral integral in the region between 500-3150 $cm^{-1}$ like we did for the experimental spectra.

A detailed derivation of our method can be found in SI chapters S7 and S8. The following gives a brief overview over the fundamental concepts and working equations.

<u>(Atomic) Raman tensors and (atomic) Raman intensities</u>

The Raman tensor $\boldsymbol{R}_k$ describes how the normal mode $k$ modulates the polarizability of a given system which determines the amplitude of the Raman-scattered light. It is given by the derivative of the system's static polarizability $\boldsymbol{\alpha}$ with respect to the displacement of the normal mode $Q_k$.

$$R_{ij} = \frac{\partial \alpha_{ij}}{\partial Q_k} \tag{1}$$

A normal mode can be understood as combination of atomic motions (see e.g. [22]). With that, this global Raman tensor can be decomposed into atomic contributions.

$$R_{ij} = \frac{\partial \alpha_{ij}}{\partial Q_k} = \sum_{n}^{atoms} \frac{1}{\sqrt{\mu_k}} \sum_{l=x,y,z} \boldsymbol{\phi}_n^l \left( \frac{\partial \alpha_{ij}}{\partial \xi_l^n} \right) \tag{2}$$

$\xi_l^n$ is the motion of atom $n$ in direction $l$, $\boldsymbol{\phi}_n^l$ the corresponding normalized atomic displacement and $\mu_k$ the normal mode's reduced trace. As the first sum goes over all atoms, the single summands are defined as atomic Raman tensors $R_{ij}^n$. The Raman intensity can be calculated from the global Raman tensor's isotropic and anisotropic contributions. Consequently, we define isotropic and anisotropic atomic contributions from which we define atomic Raman intensities. The isotropic contribution of the global Raman tensor is given by the tensor's reduced trace $\bar{\alpha}_k$.

$$\bar{\alpha}_k = \frac{1}{3}\left(R_{xx} + R_{yy} + R_{zz}\right) \tag{3}$$

Analogously, atomic reduced traces $\bar{\alpha}_k^n$ can be defined from the atomic Raman tensors. The bilinear product

$$a_k^n = \bar{\alpha}_k^n \bar{\alpha}_k \tag{4}$$

projects the atomic reduced trace onto the global reduced trace which describes an atom's isotropic contribution to the global Raman intensity. The atomic anisotropic contribution $\gamma_k^n$ is defined in analogy to the global anisotropy parameter again via a bilinear mapping of atomic and global Raman tensor entries.

$$\begin{aligned}\gamma_k^n = \frac{1}{2}\big[&\left(R_{xx}^n - R_{yy}^n\right)\left(R_{xx} - R_{yy}\right) + \left(R_{yy}^n - R_{zz}^n\right)\left(R_{yy} - R_{zz}\right) + (R_{zz}^n - R_{xx}^n)(R_{zz} - R_{xx})\big] \\ &+ 3\left[R_{xy}^n R_{xy} + R_{xz}^n R_{xz} + R_{yz}^n R_{yz}\right]\end{aligned} \tag{5}$$

From the atomic isotropic and anisotropic contributions, the atomic Raman activity $S_k^n$ can be defined as

$$S_k^n = 45 a_k^n + 7\gamma_k^n \,. \tag{6}$$

All atomic Raman activities sum up to the global Raman activity (as we prove in SI chapter S7) so that a Raman activity spectrum can be decomposed into significant atomic contributions. Finally, atomic Raman intensities can be defined from these atomic Raman activities and a mode-dependent prefactor.

$$I_k^n = \frac{(2\pi)^4}{45}(\nu_k - \nu_L)^4 \frac{h}{8\pi^2 c \nu_k \left(1 - exp\left(-\frac{h\nu_k c}{kT}\right)\right)} \cdot S_k^n \tag{7}$$

$\nu_k$ is hereby the normal mode frequency, $\nu_L$ the laser frequency, $h$ Planck's constant, $c$ the speed of light and $T = 298\,K$. The sum of these atomic Raman intensities gives the global Raman intensity as the sum over the atomic Raman activities give the global Raman activity.

Atomic Raman Polarizability Densities (RPDs) and Charge Density Differences (CDDs)

We define Raman Polarizability Densities (RPD) similar to Chen and Liu[33] and Liu *et al.*[37] over the second derivative of the molecular electron density $\rho(\boldsymbol{r})$ with the electric field $E_j$ and the normal mode displacement $Q_k$ (more details in chapter S8). $\boldsymbol{r}$ denotes here the position vector in cartesian coordinates which will later be evaluated numerically on a defined grid.

$$\frac{\partial \alpha_{ij}}{\partial Q_k} = -\int \boldsymbol{r}_i \frac{\partial^2 \rho(\boldsymbol{r})}{\partial E_j \partial Q_k} d\boldsymbol{r} \tag{8}$$

with

$$\chi_j(\boldsymbol{r}) \equiv \frac{\partial^2 \rho(\boldsymbol{r})}{\partial E_j \partial Q_k}. \tag{9}$$

We define $\chi_j(\boldsymbol{r})$ like Chen and Liu via finite element method (FEM) differentiation as response of the electron density to an electric field $E_j^{\pm}$ along the normal mode $Q_{\pm} = Q_0 \pm \Delta Q_k \boldsymbol{\varphi}_k$. Distorted molecular geometries $Q_{\pm}$ are obtained by adding/subtracting the real normal mode displacement vector $\boldsymbol{\varphi}_k$ to/from the ground state geometry $Q_0$. The real normal mode displacement vector $\boldsymbol{\varphi}_k$ is obtained from the normalized normal mode displacement vector $\boldsymbol{\phi}_k$ Gaussian16 prints out upon denormalization (see chapter S8 for more details).

$$\chi_j(\boldsymbol{r}) = \frac{\left(\rho_{Q_+}^{E_j^+} - \rho_{Q_+}^{E_j^-}\right) - (\rho_{Q_-}^{E_j^+} - \rho_{Q_-}^{E_j^-})}{4\Delta Q_k \Delta E_j} \tag{10}$$

Good FEM performance is achieved by a small differentiation grid, i.e. $\Delta E_j = 0.001$ a.u. and $\Delta Q_k = 0.05$. The volume integral over $\chi_j(\boldsymbol{r})$ must yield 0 as

$$\int \chi_j(\boldsymbol{r})\, d\boldsymbol{r} = \frac{\partial^2}{\partial E_j \partial Q_k} \int \rho(\boldsymbol{r})\, d\boldsymbol{r} = \frac{\partial^2 N}{\partial E_j \partial Q_k} = 0 \tag{11}$$

(with $N$ being the total number of electrons). We find that the $\chi_j(\boldsymbol{r})$ obtained by eq. (10) do not suffice eq. (11). We introduce a correction term that corrects $\chi_j(\boldsymbol{r})$ for that background charge noise yielding $\chi_j'(\boldsymbol{r})$ now sufficing eq. (11).

$$\chi_j'(\boldsymbol{r}) = \chi_j(\boldsymbol{r}) - \frac{1}{V} \int \chi_j(\boldsymbol{r})\, d\boldsymbol{r} \tag{12}$$

We define now the Raman Polarizability Density $R_{ij}(\boldsymbol{r})$ as

$$R_{ij}(\boldsymbol{r}) = \boldsymbol{r}_i \chi_j'(\boldsymbol{r}) \tag{13}$$

whose integral gives the Raman tensor entry

$$\frac{\partial \alpha_{ij}}{\partial Q_k} = -\int R_{ij}(\boldsymbol{r})\, d\boldsymbol{r} \tag{14}$$

(with slight numerical noise). In analogy to our definition of atomic Raman intensities, we now define first a reduced trace density $\bar{R}(\boldsymbol{r})$

$$\bar{R}(\boldsymbol{r}) = \frac{1}{3}\Big(R_{xx}(\boldsymbol{r}) + R_{yy}(\boldsymbol{r}) + R_{zz}(\boldsymbol{r})\Big) \tag{15}$$

from which the isotropic RPD contribution

$$R_{iso}(\boldsymbol{r}) = \bar{R}(\boldsymbol{r})\bar{\alpha}_k \tag{16}$$

can be defined. The anisotropic RPD is analogously

$$\begin{aligned} R_{aniso}(\boldsymbol{r}) = \frac{1}{2}\Big[&\Big(R_{xx}(\boldsymbol{r}) - R_{yy}(\boldsymbol{r})\Big)\left(R_{xx} - R_{yy}\right) + \Big(R_{yy}(\boldsymbol{r}) - R_{zz}(\boldsymbol{r})\Big)\left(R_{yy} - R_{zz}\right) \\ &+ \left(R_{zz}(\boldsymbol{r}) - R_{xx}(\boldsymbol{r})\right)\left(R_{zz} - R_{xx}\right)\Big] + 3\left[R_{xy}(\boldsymbol{r})R_{xy} + R_{xz}(\boldsymbol{r})R_{xz} + R_{yz}(\boldsymbol{r})R_{yz}\right]. \end{aligned} \tag{17}$$

The Raman activity density is given by

$$S_k(\boldsymbol{r}) = 45R_{iso}(\boldsymbol{r}) + 7R_{aniso}(\boldsymbol{r}) \tag{18}$$

that integrates to the global Raman activity (proven in chapter S8). It can be converted with the mode specific factor to the Raman Intensity Density, RID.

$$I_k(\boldsymbol{r}) = \frac{(2\pi)^4}{45}(\nu_k - \nu_L)^4 \frac{h}{8\pi^2 c\nu_k \left(1 - exp\left(-\frac{h\nu_k c}{kT}\right)\right)} S_k(\boldsymbol{r}) \tag{19}$$

We define the molecular plane as xz-plane so that we can neglect all y-contribution to lower the computational effort (see SI chapter S8 for the reduced working equations). Atomic RIDs can now be analogously defined by defining single atomic displaced geometries. Instead of adding/subtracting the whole displacement vector $\boldsymbol{\varphi}_k$, only the three (x, y and z) entries of atom $n$ are added/subtracted from the molecule's ground state geometry. FEM differentiation then yields again the atomic $\chi_j'(\boldsymbol{r})$ from eq. (12) and the sum over all atomic RIDs yields the global RID for each normal mode. Furthermore, the atomic RIDs integrate to the atomic Raman intensities. The RIDs shown in this paper are obtained by projecting the calculated (atomic) $I_k(\boldsymbol{r})$ maps onto the electron densities of the molecules' ground state geometries (with a fixed isodensity value of 0.005 a.u. $a_0^{-3}$) for clarity. This projection results in a reduction in information; however, it allows for a chemically intuitive interpretation. (Atomic) CDDs are analogously defined from difference in electrostatic potentials (ESPs) in the (single atomic) displaced geometries. They are also projected onto the ground state geometries for clarity.

## Acknowledgements:

M.B. is grateful to Evangelisches Studienwerk e.V. Villigst for financial support. The authors acknowledge support by the state of Baden-Württemberg through bwHPC and the German Research

Foundation (DFG) through grant no INST 40/575- 1 FUGG (JUSTUS 2 cluster). The work in Madrid was supported by grant PID2022-138222NB-C21 funded by MICIU/AEI/10.13039/501100011033 and by ERDF/EU, and by the Severo Ochoa program for Centers of Excellence in R&D; grant no. CEX2025-001660-S.

**Competing interests:**

The authors declare no competing interests.

**Data availability:**

Additional info as well as molecular structures are presented in the SI. All scripts used for computing atomic Raman intensities and RIDs will have been published upon publication of this article.

# Supplementary Info: Visualizing and Quantifying Atomic Contributions to Raman Intensities governed by Spatially-Resolved Atomic Interferences

*Marc Bröckel[1,2], Johannes Gierschner[1,3], Alfred J. Meixner[1,2] and Kai Braun[1,2]**

[1] *Institute of Physical and Theoretical Chemistry, University of Tübingen, Auf der Morgenstelle 18, 72076 Tübingen, Germany*

[2] *Center for Light-Matter Interaction, Sensors&Analytics LISA+, University of Tübingen, Auf der Morgenstelle 15, 72076 Tübingen, Germany*

[3] *Madrid Institute for Advanced Studies, IMDEA Nanoscience, Madrid, Spain*

---

**Contents**

## S1. Band Assignment of MBT, CMBT and EMBT

| MBT | | | CMBT | | | EMBT | | | Assignment | | |
|---|---|---|---|---|---|---|---|---|---|---|---|
| $\nu_{exp}$ / $cm^{-1}$ | $\nu_{calc}$ / $cm^{-1}$ | $0.954 \cdot \nu_{calc}$ / $cm^{-1}$ | $\nu_{exp}$ / $cm^{-1}$ | $\nu_{calc}$ / $cm^{-1}$ | $0.954 \cdot \nu_{calc}$ / $cm^{-1}$ | $\nu_{exp}$ / $cm^{-1}$ | $\nu_{calc}$ / $cm^{-1}$ | $0.954 \cdot \nu_{calc}$ / $cm^{-1}$ | | | |
| | | | 409 | 400 | 382 | | | | δ | ip | S16-C15-S17 |
| 402 | 413 | 394 | 558 | 599 | 571 | | | | δ | ip | C13-C12-S16 |
| 459 | 450 | 429 | | | | | | | δ | oop | |
| 523 | 523 | 499 | 473 | 476 | 454 | 571 | 555 | 529 | δ | ip | C12-S16-C15 |
| | | | 600 | | | | | | | | |
| 607 | 627 | 598 | 628 | 647 | 617 | 648 | 605 | 577 | δ | ip | C8-C11 stretch |
| | 629 | 600 | | | | | | | δ | oop | |
| 730 | 725 | 692 | 726 | 734 | 700 | 724 | 723 | 690 | ν | as | C12-S16-C15 |
| | 745 | 711 | 760 | 776 | 740 | | 735 | 701 | ν | s | C12-S16-C15 |
| 771 | 765 | 730 | | 770 | 735 | | | | δ | oop | benzene |
| | | | 836 | 844 | 805 | | | | δ | oop | benzene |
| 874 | 894 | 853 | 932 | 947 | 903 | 840 | 861 | 821 | ν | | thiazole breathing |
| | | | | | | 901 | 917 | 875 | δ | ip | benzene breathing + Et |
| | | | | | | 948 | 977 | 932 | δ | ip | benzene breathing + Et |
| 1022 | 1049 | 1001 | 1026 | 1050 | 1002 | 1022 | 1049 | 1001 | ν | s | C12-S16-C15 |
| 1022 | 1062 | 1013 | 1093 | 1113 | 1062 | 1062 | 1086 | 1036 | δ | ip | benzene |
| 1070 | 1121 | 1069 | 1090 | 1123 | 1071 | 1100 | 1114 | 1062 | ν | as | S16-C15-S17 |
| 1147 | 1167 | 1113 | | | | 1141 | 1156 | 1103 | δ | ip | benzene C9H19-C10H20 |
| | 1195 | 1140 | 1165 | 1187 | 1132 | | 1161 | 1108 | δ | ip | benzene C8H18-C13H21 |
| 1258 | 1294 | 1234 | 1251 | 1295 | 1235 | 1232 | 1276 | 1217 | ν | | C11-N14 |
| | | | | | | 1278 | 1309 | 1249 | δ | ip | benzene C-H |
| 1297 | 1327 | 1266 | 1251 | 1287 | 1228 | | 1322 | 1261 | ν | | C11-N14 |
| | 1348 | 1286 | 1329 | 1334 | 1273 | 1336 | 1361 | 1298 | ν | | benzene |
| 1406 | 1477 | 1409 | 1406 | 1471 | 1403 | 1406 | 1469 | 1401 | ν | | N14-C15 |
| 1406 | 1514 | 1444 | 1406 | 1475 | 1407 | 1465 | 1529 | 1459 | ν | s | C11-N14-C15 |
| | | | | | | 1444 | 1492 | 1423 | δ | oop | twist Et |
| 1463 | 1519 | 1449 | 1463 | 1517 | 1447 | 1444 | 1504 | 1435 | ν | as | C11-N14-C15 |
| 1595 | 1646 | 1570 | 1571 | 1633 | 1558 | 1570 | 1642 | 1566 | ν | | benzene |
| 1595 | 1675 | 1598 | 1602 | 1670 | 1593 | 1614 | 1684 | 1607 | ν | | benzene |
| | | | | | | 2889 | 3023 | 2884 | ν | s | methylene |
| | | | | | | 2942 | 3061 | 2920 | ν | as | methylene |
| 3078 | 3197 | 3050 | | | | | | | ν | | benzene C-H |
| | 3204 | 3057 | 3100 | 3213 | 3065 | 3078 | 3203 | 3056 | ν | | benzene C-H |
| | 3214 | 3066 | | 3221 | 3073 | | 3225 | 3077 | ν | | benzene C-H |
| | 3223 | 3075 | | 3232 | 3083 | | 3242 | 3093 | ν | | benzene C-H |

**S2. <u>Normal mode symmetries</u>**

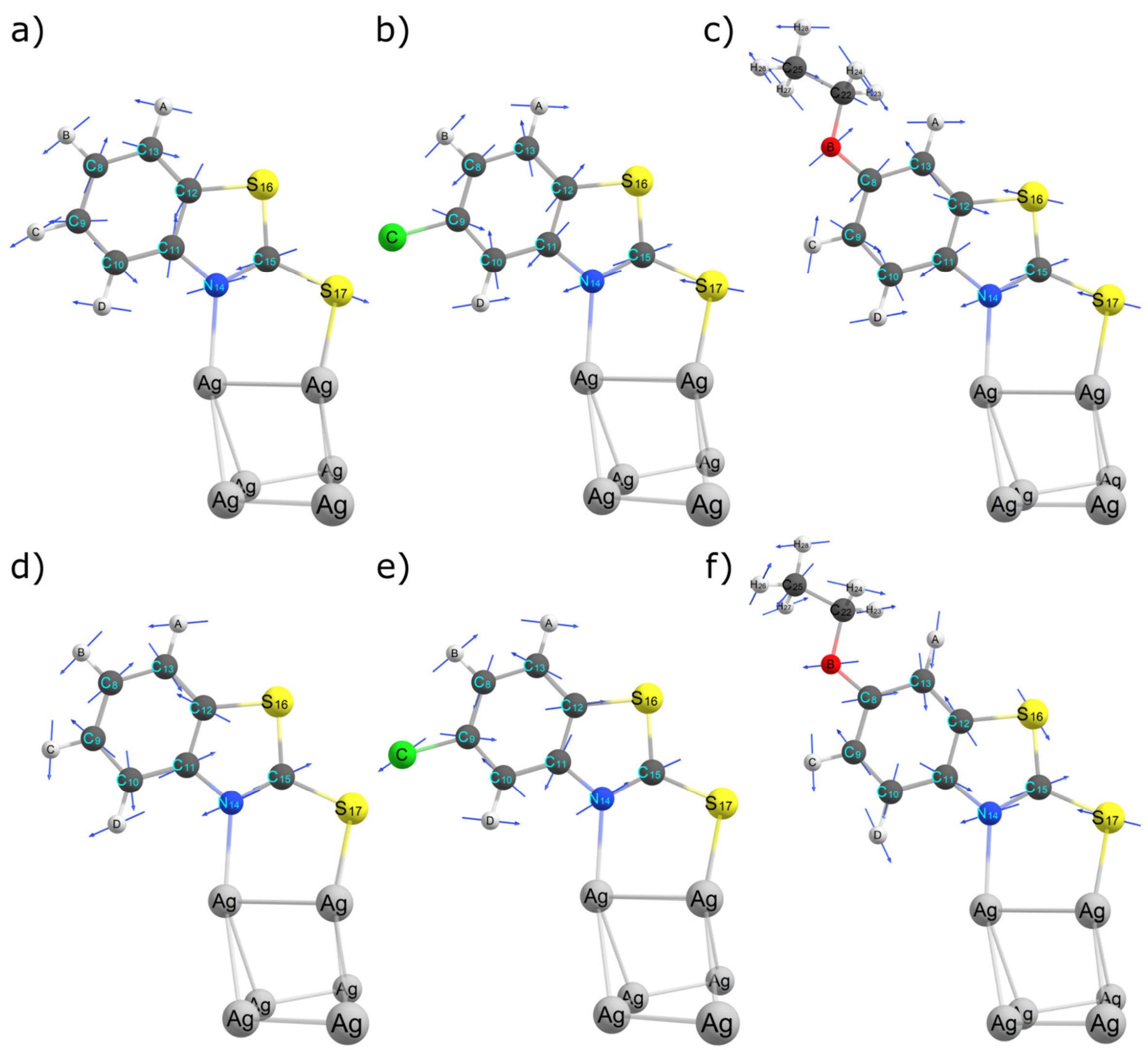

*Figure S1: Displacement vector directions of the 1406 cm$^{-1}$ normal mode of a) MBT, b) CMBT and c) EMBT and d) – f) the 1463 cm$^{-1}$ mode of the three molecules.*

We see in Figure S1 that the 1406 cm$^{-1}$ normal modes of MBT and CMBT are equivalent while the C12 motion changes in EMBT due to the shift of the center of mass. Similarly, the C11 and C12 motions change in CMBT and EMBT for the 1463 cm$^{-1}$ mode which is represented in the atomic Raman intensities discussed in the main text.

**S3. Electronic absorption spectra of the MBT, CMBT and EMBT**

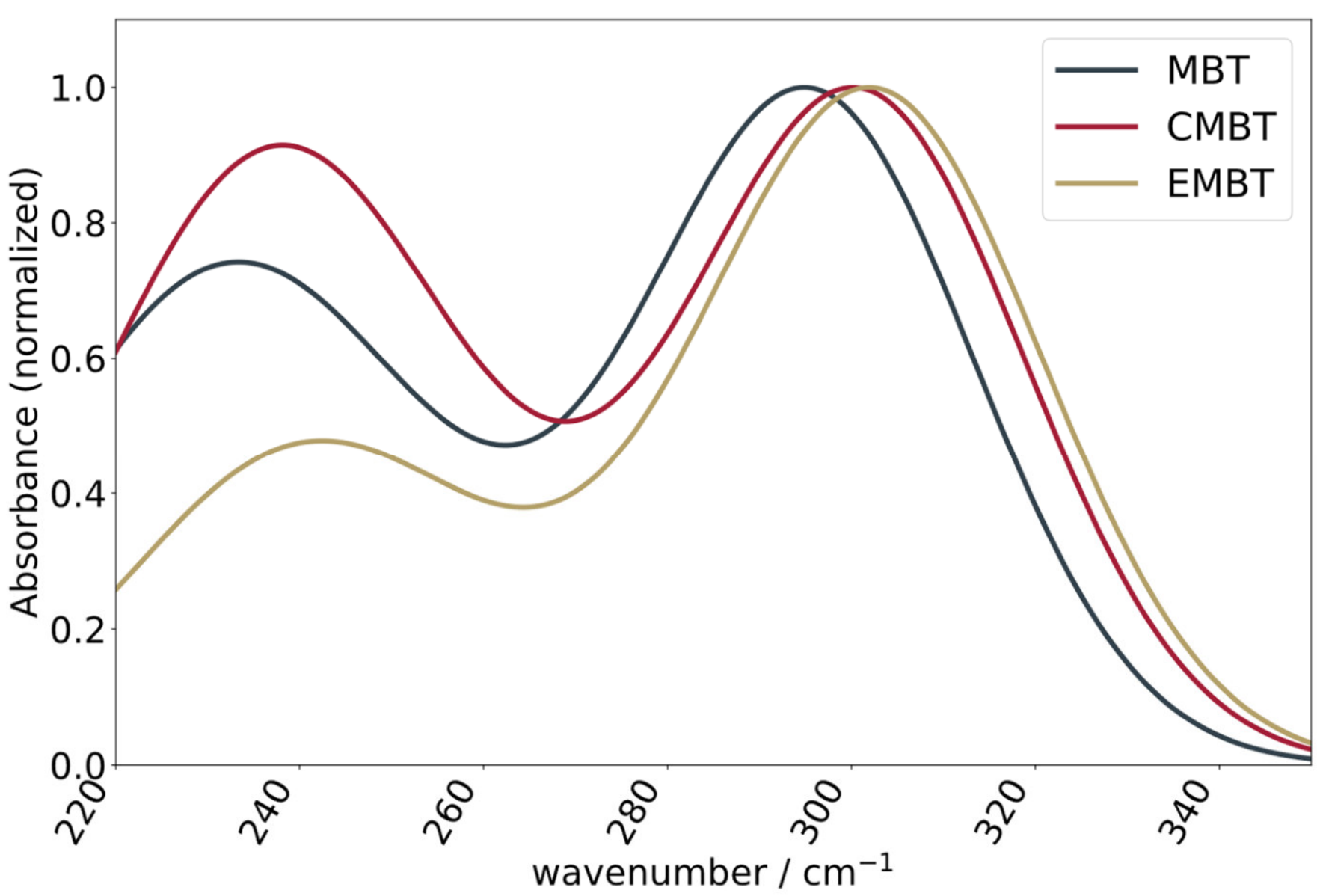

*Figure S2: Electronic absorption spectra of MBT, CMBT and EMBT obtained at the TD- B3LYP/def2-TZVP level of theory.*

Figure S2 shows the calculated electronic absorption spectra of MBT, CMBT and EMBT (single molecules, not the adsorbates) as obtained by TDDFT (B3LYP[1,2]/def2-TZVPP[3,4]; spectra were obtained by convoluting a gaussian with a FWHM of 30 nm). All reported absorption bands are significantly higher in energy than the 532 nm laser used for SERS experiments so that resonance Raman effects may be neglected in this study.

**S4. Cluster ansatz for calculating SERS spectra**

To correctly reproduce the experimental SERS spectra, we had to find a chemical model for the SERS sample. We chose the cluster ansatz established by Zhao et al.[5] and Wu et al.[6] to incorporate the chemical effect of the nearest silver atoms to the molecules. For that, we first optimized 11 different Ag clusters consisting of 4 to 20 Ag atoms using the long-range-corrected CAM-B3LYP[7] hybrid functional corrected with Grimme's D3[8] dispersion correction and Becke-Johnson damping[9]. We used the Stuttgart-Dresden (SDD) MWB28[10,11] effective core potential (ECP) for the Ag atoms. All clusters were treated as closed-shell systems so that for uneven numbers of Ag atoms either a positive or negative charge was assigned to the clusters. Frequency calculations confirmed local minima in the clusters' potential energy surfaces (PESs). The resulting stable cluster geometries are shown in Figure S3.

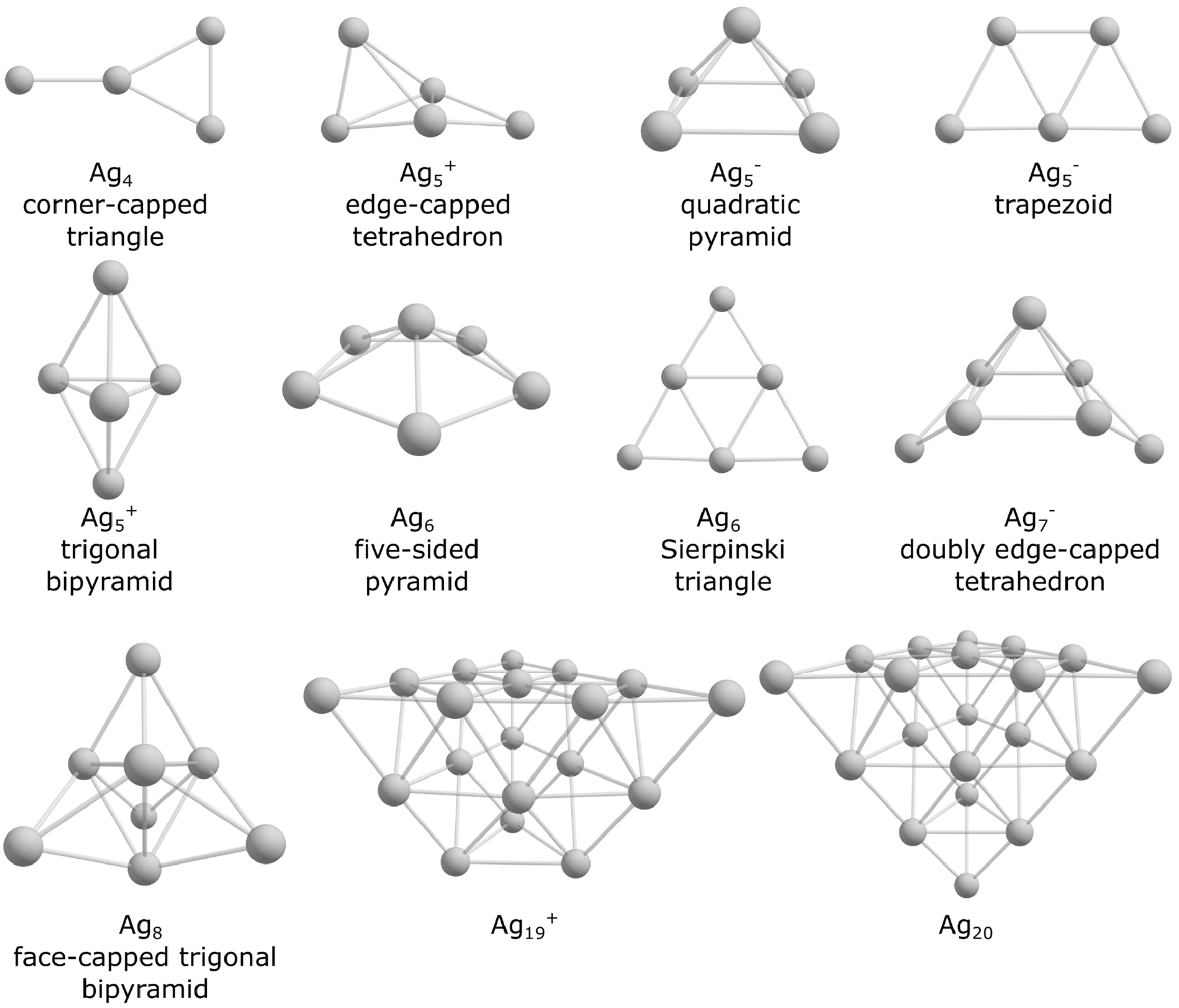

*Figure S3: Different stable Ag cluster geometries to model the chemical influence of the silver surface in SERS calculations.*

Subsequently, a preoptimized molecular structure of MBT, CMBT or EMBT (CAM-B3LYP-D3BJ/def2-TZVPP) was added to these cluster geometries and adsorption geometries were optimized (CAM-B3LYP-D3BJ/SDD(Ag),cc-pVTZ(H,C,N,O,Cl)). Frequency calculations again confirmed minima in the adsorbates' PESs and were used to calculate Raman spectra. The considered adsorbates are shown in Figure S4. We then calculated Raman spectra where all harmonic frequencies are scaled by a scaling factor of 0.954[12] and broadened with a Gaussian with a FWHM of 20 $cm^{-1}$. We normalized all calculated spectra to the spectral integral in the region between 500-3150 $cm^{-1}$ like we did for the experimental spectra. We found the best agreement between the experimental SERS spectrum and the (S,N)-$Ag_7$ doubly edge-capped tetrahedron. Therefore, we chose this geometry (and the corresponding geometries for CMBT and EMBT) for all further computational consideration. They are as follows (coordinates in a.u.).

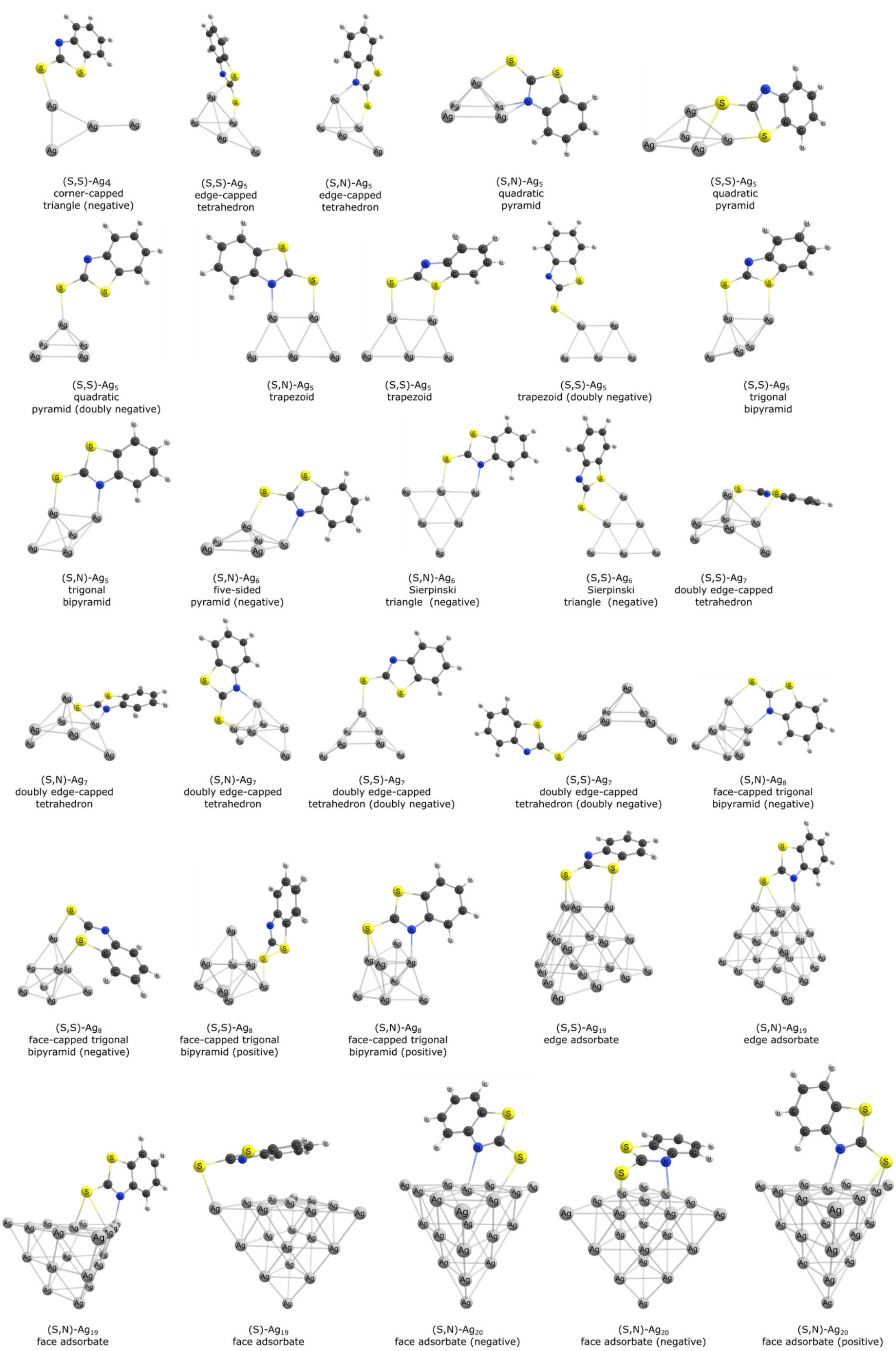

*Figure S4: Studied stable (C/E)MBT-Ag adsorbates for calculating SERS spectra.*

MBT-$Ag_7$-adsorbate

| | | | |
|---|---|---|---|
| Ag | 1.721947000000 | -1.356962000000 | 1.426144000000 |
| Ag | 2.357536000000 | -3.001627000000 | -0.577846000000 |
| Ag | 0.194946000000 | 1.324774000000 | -1.174093000000 |
| Ag | 2.158328000000 | 3.125199000000 | -0.537565000000 |
| Ag | 0.203509000000 | -1.316769000000 | -1.125198000000 |
| Ag | 1.563705000000 | 1.384217000000 | 1.486271000000 |
| Ag | -0.784119000000 | -0.051799000000 | 1.062910000000 |
| C | -6.622473000000 | -0.082470000000 | 1.587040000000 |
| C | -5.554646000000 | -0.103403000000 | 2.484180000000 |
| C | -4.251803000000 | -0.100693000000 | 2.035884000000 |
| C | -4.019046000000 | -0.076684000000 | 0.666747000000 |
| C | -5.092916000000 | -0.055669000000 | -0.225482000000 |
| C | -6.403705000000 | -0.058401000000 | 0.225086000000 |
| N | -2.773994000000 | -0.070976000000 | 0.059498000000 |
| C | -2.852109000000 | -0.046530000000 | -1.237805000000 |
| S | -4.481174000000 | -0.027917000000 | -1.854105000000 |
| S | -1.538147000000 | -0.036145000000 | -2.379883000000 |
| H | -7.635298000000 | -0.085023000000 | 1.962883000000 |
| H | -5.752787000000 | -0.121946000000 | 3.546020000000 |
| H | -3.418568000000 | -0.116647000000 | 2.724356000000 |
| H | -7.231088000000 | -0.042043000000 | -0.469098000000 |

CMBT-$Ag_7$-adsorbate

| | | | |
|---|---|---|---|
| Ag | -1.675964000000 | -1.392005000000 | -1.570467000000 |
| Ag | -2.644745000000 | -3.010651000000 | 0.318005000000 |
| Ag | -0.652885000000 | 1.346464000000 | 1.211404000000 |
| Ag | -2.485255000000 | 3.116442000000 | 0.205845000000 |
| Ag | -0.633814000000 | -1.295486000000 | 1.208241000000 |
| Ag | -1.529127000000 | 1.349317000000 | -1.649848000000 |
| Ag | 0.716753000000 | -0.055850000000 | -0.792423000000 |
| C | 6.562300000000 | -0.033850000000 | -0.263833000000 |
| C | 5.661259000000 | -0.079822000000 | -1.325287000000 |
| C | 4.299896000000 | -0.084021000000 | -1.127528000000 |
| C | 3.833437000000 | -0.040620000000 | 0.179713000000 |
| C | 4.730292000000 | 0.005742000000 | 1.248800000000 |
| C | 6.099074000000 | 0.009354000000 | 1.033765000000 |
| N | 2.501010000000 | -0.038442000000 | 0.551906000000 |
| C | 2.345711000000 | 0.007387000000 | 1.843058000000 |
| S | 3.838023000000 | 0.052826000000 | 2.740562000000 |
| S | 0.850794000000 | 0.022273000000 | 2.730955000000 |
| H | 7.621796000000 | -0.032350000000 | -0.467603000000 |
| H | 3.612821000000 | -0.119719000000 | -1.959560000000 |
| H | 6.794461000000 | 0.045210000000 | 1.859259000000 |
| Cl | 6.281626000000 | -0.133384000000 | -2.948081000000 |

EMBT-$Ag_7$-adsorbate

| | | | |
|---|---|---|---|
| Ag | -2.066266000000 | -1.367192000000 | -1.548097000000 |
| Ag | -2.903816000000 | -3.009901000000 | 0.381900000000 |
| Ag | -0.836037000000 | 1.328140000000 | 1.191684000000 |
| Ag | -2.728676000000 | 3.117681000000 | 0.344909000000 |
| Ag | -0.827624000000 | -1.313526000000 | 1.149957000000 |
| Ag | -1.914711000000 | 1.374548000000 | -1.599173000000 |
| Ag | 0.381215000000 | -0.049295000000 | -0.924604000000 |
| C | 6.256397000000 | -0.049814000000 | -0.820697000000 |
| C | 5.282752000000 | -0.078998000000 | -1.830710000000 |
| C | 3.946599000000 | -0.082445000000 | -1.525973000000 |
| C | 3.557547000000 | -0.056488000000 | -0.189421000000 |
| C | 4.529400000000 | -0.027499000000 | 0.805406000000 |
| C | 5.887950000000 | -0.023667000000 | 0.512488000000 |
| N | 2.253782000000 | -0.056087000000 | 0.280678000000 |
| C | 2.191454000000 | -0.027953000000 | 1.576635000000 |
| S | 3.749781000000 | 0.000985000000 | 2.359608000000 |
| S | 0.763352000000 | -0.021513000000 | 2.576553000000 |
| H | 3.198952000000 | -0.104752000000 | -2.306410000000 |
| H | 6.619142000000 | -0.000844000000 | 1.303343000000 |
| H | 5.621142000000 | -0.098562000000 | -2.855808000000 |
| O | 7.536334000000 | -0.049527000000 | -1.253452000000 |
| C | 8.582308000000 | -0.020346000000 | -0.295648000000 |
| H | 8.503021000000 | -0.892087000000 | 0.359801000000 |
| H | 8.489062000000 | 0.877860000000 | 0.321082000000 |
| C | 9.889175000000 | -0.026447000000 | -1.045266000000 |

**S5. <u>Adsorption geometry of MBT on Ag</u>**

The chemisorption geometry of MBT on a Ag surface has been discussed heatedly in the last decades and is not yet unambiguously known. Ohsawa and Suetaka[13] first studied the adsorption of MBT on Cu in 1979 and proposed a layer of $Cu^{I}$MBT polymers that covers the surface. Sandhyarani et al.[14] conducted SERS and XPS experiments in 1999 to study the adsorption geometry of MBT on Au and Ag. They conclude that MBT adsorbs perpendicularly to a Au surface in its thione form but parallel to a Ag surface in its thiol form. They report a S-H vibration at 2641 $cm^{-1}$ which we can also observe in some single but not in the majority of the recorded SERS spectra. Furthermore, they neglect the possibility of binding via the thiazole N atom. Woods et al.[15] conducted similar SERS and XPS measurements of MBT on Ag in the year 2000 and they don't report the 2641 $cm^{-1}$ S-H-peak. They conclude an oxidative charge transfer adsorption of MBT via the exocyclic S atom (thus in its thione form) but they do not further investigate the adsorption geometry. Lee et al.[16] also report SERS spectra of MBT on Ag, don't observe the 2641 $cm^{-1}$ peak and also conclude from the Raman intensities that the molecule cannot adsorb flatly but has to be tilted to the surface with some respect. However, they also do not study whether MBT adsorbs via the endocyclic S or N and just propose that it adsorbs with the endocyclic S. Yang et al.[17] conducted electrochemical impedance spectroscopy as well as SERS measurements of MBT monolayers on Ag and Zn and conclude from their voltage dependent SERS spectra (in solution) that MBT adsorbs via its exocyclic S and endocyclic N on Ag. However, they also only argue with the changes in Raman intensities, which we have shown in this paper, does not necessarily stem from a change in adsorption geometry but can also be attributed to electron density shifts due to the electrical field in their spectroelectrochemical experiments. Beattie et al.[18] then conducted a thorough synchrotron XPS study of MBT on Au and Au:Ag alloys in 2009. They report two signals in the N 1s spectrum from which they conclude two adsorption geometries on Au: one they call "upright" with MBT only binding via its exocyclic S atom and a “tilted” geometry where MBT binds with both its exocyclic S and N atoms.

So, there was and still is a big controversy on the adsorption geometry of MBT on Ag. Most studies agree that MBT cannot lie flatly on the surface as that would yield high intensity out-of-plane modes which we do not see in the experiment. Most studies also agree that MBT adsorbs via its exocyclic S atom in its thione form. However, the question whether it also adsorbs via its endocyclic thiazole S or N remains. Beattie et al. have nicely shown that it must adsorb via its N atom on Au and we transfer this to adsorption on Ag as our calculations also suggest adsorption via the endocyclic N.

Directly comparing the adsorbate energies of the (S,S)- and (S,N)-$Ag_7$ doubly edge-capped tetrahedron structures reveals that the (S,S)-adsorption has a significantly higher Gibb’s free energy (81 kJ/mol). Furthermore, we report that the calculated Raman spectra of the (S,N)-adsorbate describes the experimental spectrum significantly better than all studied (S,S)-adsorbates. Figure S5 compares the experimental SERS spectrum, the calculated spectrum of the (S,N)-adsorbate reported in the previous chapter and a linear combination of seven (S,S)-adsorbate structures.

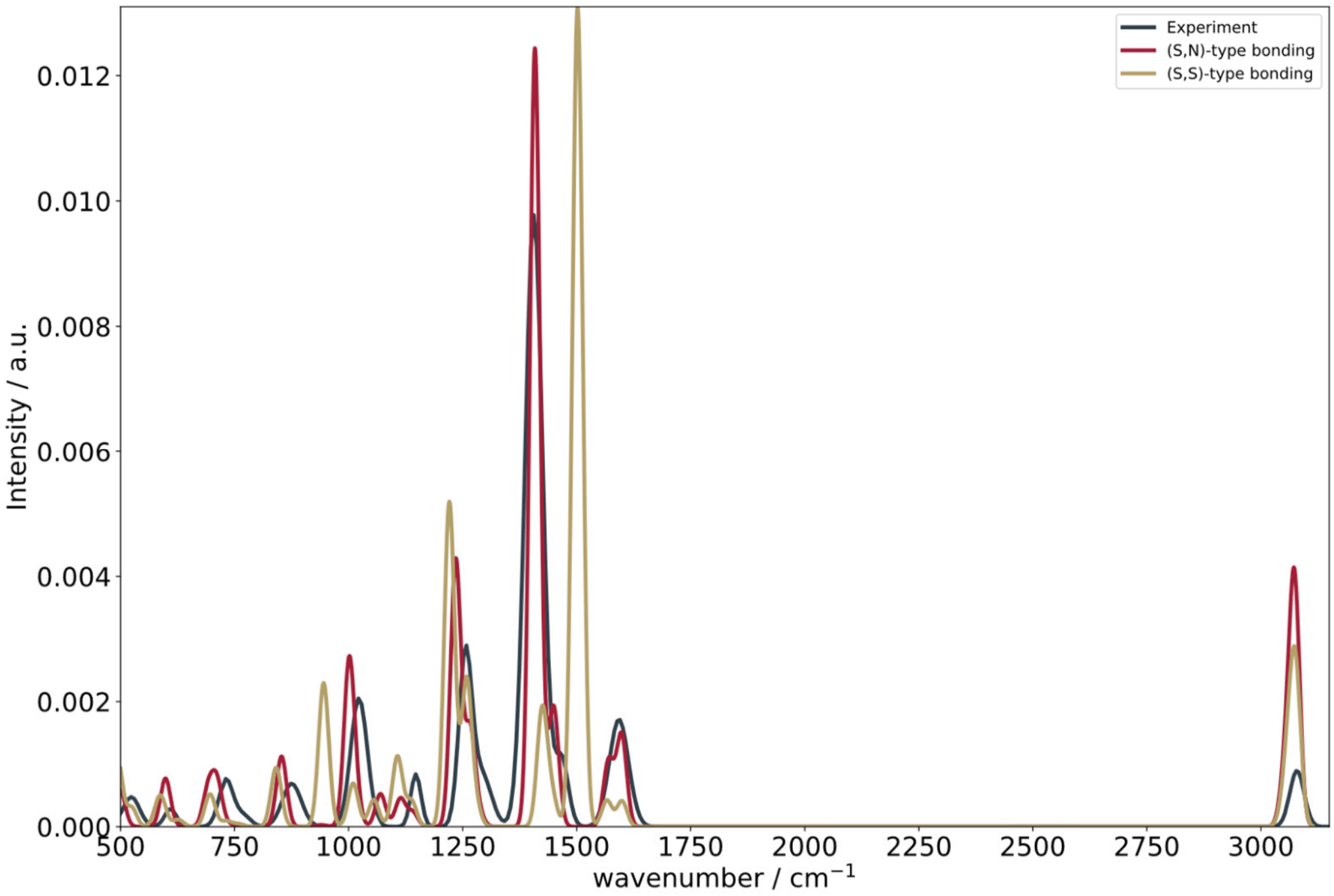

*Figure S5: Comparison of (S,N)- and (S,S)-type MBT adsorbate SERS spectra.*

The SERS spectra of both bonding types show two dominant features in the region between 1350 $cm^{-1}$ and 1500 $cm^{-1}$. Both peaks can be attributed to two similar C-N-stretching modes, and the underlying normal modes are the same normal modes for all considered adsorbates. We observe a main peak at 1406 $cm^{-1}$ and a second peak at 1463 $cm^{-1}$ with a significantly lower Raman intensity for the (S,N)-type binding. For the (S,S)-type bonding style, however, the relative intensity is reversed. The second peak at ~1500 $cm^{-1}$ shows the higher intensity. And we observe this trend in all considered (S,S)-type bonding adsorbates which does not fit the experiment. Therefore, we conclude that (S,N)-type binding better describes the molecular adsorption geometry in MBT SAMs.

**S6. <u>5-EMBT and 6-CMBT</u>**

We study 5-CMBT and 6-EMBT in this work as both molecules are easily commercially available. However, by doing so, we mix electronic and structural influences of the substituents. Here, we shortly discuss the electronic structures of 6-CMBT and 5-EMBT by means of electrostatic potential (ESP) and charge density difference (CDD) maps (with respect to MBT).

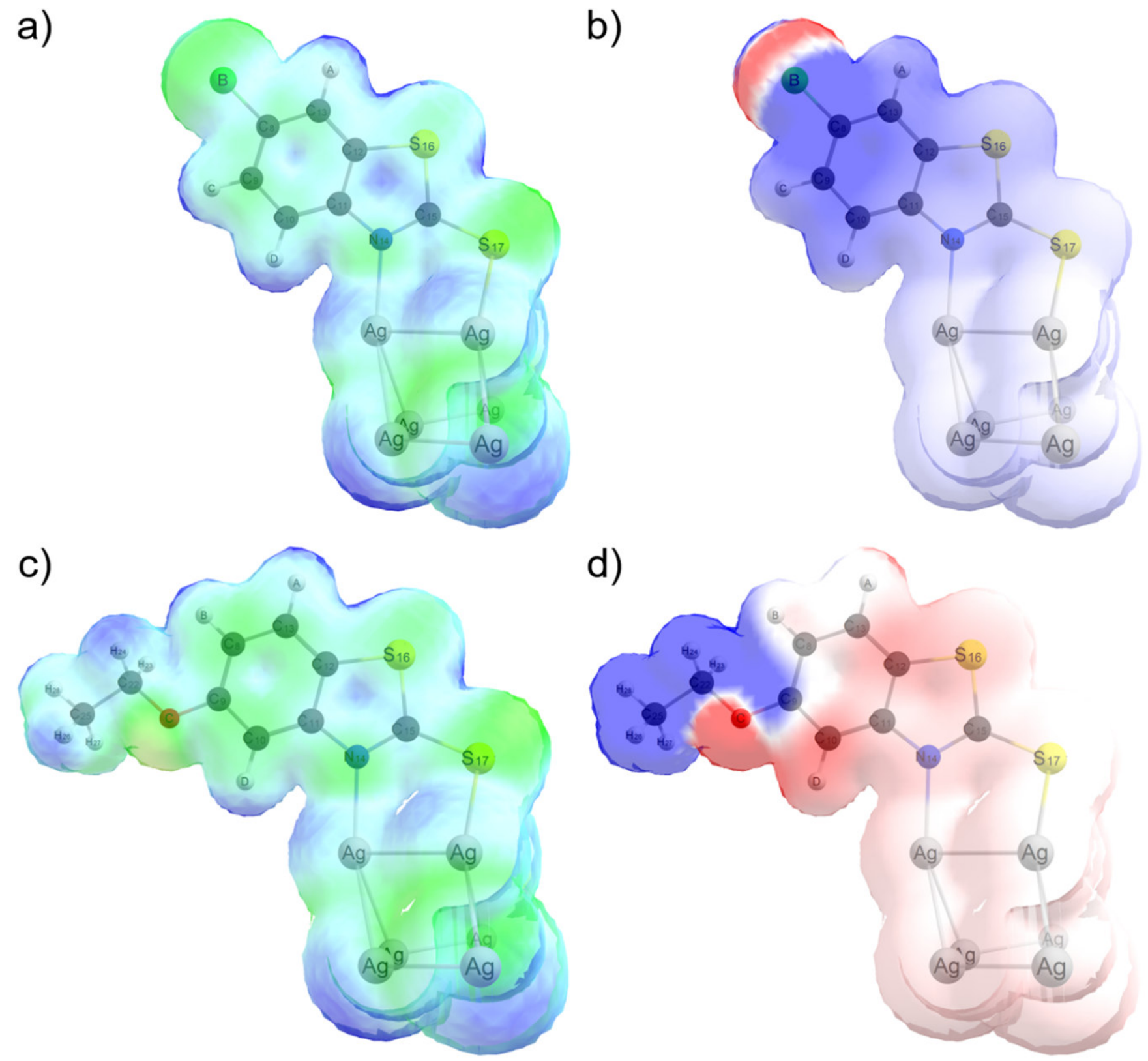

*Figure S6: ESPs of a) 6-CMBT and c) 5-EMBT and their CDDs with respect to MBT b) and d).*

6-CMBT shows, like 5-CMBT, mainly the inductive effect which withdraws electron density from the MBT basis only limited to distance expanding into the $Ag_7$-cluster. The strongest effect is experienced by the bonded C atom which is C9 for 5-CMBT and C8 for 6-CMBT. More charge density is withdrawn from S16 than S17 in both 5-CMBT and 6-CMBT, however, the amount of charge density withdrawn from the S atoms as well as the Ag-cluster is comparable. Substitution in 5- or 6-position thus does not significantly alter the electronic structure.

5-EMBT also shows a negative inductive effect mainly experienced by the bonded atom C9 and the major positive mesomeric effect with which the substituent couples to the thiazole's electron density. It is strongest in para-position and extends onto the N-C=S moiety in 6-EMBT. In 5-EMBT, it extends onto the S-C=S moiety due to the different position. This is a significant change in electronic structure. The Ag-clusters, however, appear to be similarly affected in both EMBT isomers.

**S7. <u>(Atomic) Raman tensors and intensities</u>**

Every Raman calculation precedes a harmonic frequency calculation to obtain the normal modes their corresponding frequencies. Global Raman tensors can be obtained by understanding a normal mode as a linear combination of atomic motions. Le Ru and Etchegoin[19] thoroughly discuss the properties of the Raman tensor and that Gaussian03 (the later Gaussian16 version we use uses the same mathematical concept) computes them as a linear combination of atomic displacements.

$$\boldsymbol{R}_k = \frac{\partial \boldsymbol{\alpha}}{\partial Q_k} = \sum_{n}^{atoms} \frac{1}{\sqrt{\mu_k}} \sum_{l=x,y,z} \boldsymbol{\phi}_n^l \left(\frac{\partial \boldsymbol{\alpha}}{\partial \xi_n}\right) \tag{1}$$

The global Raman tensor $\boldsymbol{R}_k$ of normal mode $k$ is defined as the derivative of the molecule's static polarizability $\alpha$ with respect to the normal mode $Q_k$. This derivative is commonly rewritten by treating the normal mode $Q_k$ as a linear combination of normalized atomic displacements $\boldsymbol{\phi}_n^l$ normalized to the reduced mass of the normal mode $\mu_k$. The index $n$ runs over all atoms and the index $l$ over the three special dimensions x, y and z. The derivative of the polarizability with a single atomic displacement $\frac{\partial \alpha}{\partial \xi_n}$ is analytically available, so that these figures are printed out in a standard Gaussian16 Raman calculation. The Raman tensor is a 3x3 tensor and its entries are referred to as $R_{ij}$ with i and j being x, y or z.

$$\boldsymbol{R}_k = \begin{pmatrix} R_{xx} & R_{xy} & R_{xz} \\ R_{yx} & R_{yy} & R_{yz} \\ R_{zx} & R_{zy} & R_{zz} \end{pmatrix} \tag{2}$$

The reduced trace of the Raman tensor is defined as

$$\bar{\alpha}_k = \frac{1}{3}\left(R_{xx} + R_{yy} + R_{zz}\right) \tag{3}$$

and the anisotropy factor as

$$\bar{\gamma}_k^2 = \frac{1}{2}\left[\left(R_{xx}\text{-}R_{yy}\right)^2 + \left(R_{yy} - R_{zz}\right)^2 + (R_{zz} - R_{xx})^2\right] + 3\left[R_{xy}^2 + R_{xz}^2 + R_{yz}^2\right]. \tag{4}$$

With this, we can calculate the Raman activity $S_k$

$$S_k = 45\bar{\alpha}_k^2 + 7\bar{\gamma}_k^2 \tag{5}$$

and the Raman intensity $I_k$

$$I_k = \frac{(2\pi)^4}{45}(\nu_k - \nu_L)^4 \cdot \frac{h}{8\pi^2 c\nu_k\left[1 - exp\left(-\frac{h\nu_k c}{kT}\right)\right]} \cdot S_k\,. \tag{6}$$

$\nu_L$ and $\nu_k$ are the wavenumbers of the excitation Laser and the normal mode, $h$ is Planck's constant, $k$ the Boltzmann constant, $c$ the speed of light and $T$ the system's temperature.

Eq. (1) shows that Gaussian16 calculates the global Raman tensor as sum over atomic Raman tensors $\boldsymbol{R}_k^n$.

$$\boldsymbol{R}_k^n = \frac{1}{\sqrt{\mu_k}} \sum_{l=x,y,z} \boldsymbol{\phi}_n^l \left(\frac{\partial \boldsymbol{\alpha}}{\partial \xi_n}\right). \tag{7}$$

From these atomic Raman tensors, we analogously define atomic reduced traces $\bar{\alpha}_k^n$ which are used in the main text to derive atomic Raman intensities. This tensor decomposition features that the sum over the atomic Raman tensor entries yields the corresponding global Raman tensor entry.

$$R_{ij} = \sum_n R_{ij}^n \tag{8}$$

With that, also the atomic reduced traces sum up to the global reduced trace. Consequently, the isotropic atomic contribution $a_k^n$ sums up to the squared global Raman tensor reduced trace.

$$\sum_n a_k^n = \sum_n \bar{\alpha}_k^n \bar{\alpha}_k = \bar{\alpha}_k^2 \tag{9}$$

Analogously, the anisotropic atomic contributions yield the global anisotropy parameter $\bar{\gamma}_k^2$

$$\begin{aligned} \sum_n \gamma_k^n &= \frac{1}{2}\sum_n \left[\left(R_{xx}^n - R_{yy}^n\right)\left(R_{xx} - R_{yy}\right) + \left(R_{yy}^n - R_{zz}^n\right)\left(R_{yy} - R_{zz}\right) + \left(R_{zz}^n - R_{xx}^n\right)\left(R_{zz} - R_{xx}\right)\right] \\ &\qquad + 3\sum_n \left[R_{xy}^n R_{xy} + R_{xz}^n R_{xz} + R_{yz}^n R_{yz}\right] \\ &= \frac{1}{2}\left[\left(R_{xx} - R_{yy}\right)\left(R_{xx} - R_{yy}\right) + \left(R_{yy} - R_{zz}\right)\left(R_{yy} - R_{zz}\right) + \left(R_{zz} - R_{xx}\right)\left(R_{zz} - R_{xx}\right)\right] \\ &\qquad + 3\left[R_{xy}R_{xy} + R_{xz}R_{xz} + R_{yz}R_{yz}\right] \\ &= \frac{1}{2}\left[\left(R_{xx}-R_{yy}\right)^2 + \left(R_{yy} - R_{zz}\right)^2 + \left(R_{zz} - R_{xx}\right)^2\right] + 3\left[R_{xy}^2 + R_{xz}^2 + R_{yz}^2\right] \\ &= \bar{\gamma}_k^2 \end{aligned} \tag{10}$$

So, with defining the atomic Raman activity as

$$S_k^n = 45 a_k^n + 7 \gamma_k^n , \tag{11}$$

we obtain a set of atomic Raman activities whose sum yields the global Raman activity.

$$\sum_n S_k^n = 45 \sum_n a_k^n + 7 \sum_n \gamma_k^n = 45\bar{\alpha}_k^2 + 7\bar{\gamma}_k^2 = S_k \tag{12}$$

Consequently, the atomic Raman intensities defined from these atomic Raman activities also add up to the global Raman intensity.

**S8. <u>(Atomic) Raman Intensity Densities (RIDs) and charge density differences (CDDs)</u>**

The main derivation of RPDs and CDDs are given in the main text. However, we would like to give some more details here. The molecular polarizability is given as

$$\alpha_{ij} = \frac{\partial \mu_i}{\partial E_j} \tag{13}$$

with the dipole moment being defined from the center of positive charges to the center of negative charges

$$\boldsymbol{\mu} = \sum_A \boldsymbol{R}_A Z_A - \int \boldsymbol{r}\rho(\boldsymbol{r})\, d\boldsymbol{r}\,. \tag{14}$$

The electron density $\rho(\boldsymbol{r})$ describes the distribution of the electrons around a given molecular geometry. Therefore, it should be denoted as $\rho(\boldsymbol{r}; \boldsymbol{R}_A, Z_A)$, however it is commonly reduced to $\rho(\boldsymbol{r})$ for clarity. The partial derivative of the electric dipole moment yields

$$\alpha_{ij} = \frac{\partial}{\partial E_j}\left(\sum_A \boldsymbol{R}_A Z_A - \int \boldsymbol{r}_i\rho(\boldsymbol{r})\, d\boldsymbol{r}\right) = -\int \boldsymbol{r}_i \frac{\partial \rho(\boldsymbol{r})}{\partial E_j} d\boldsymbol{r} \tag{15}$$

in the Born-Oppenheimer approximation as the electric field does not influence the nuclei. The Raman intensity is proportional to the change in polarizability during the normal mode yielding

$$\frac{\partial \alpha_{ij}}{\partial Q_k} = -\int \boldsymbol{r}_i \frac{\partial^2 \rho(\boldsymbol{r})}{\partial E_j \partial Q_k} d\boldsymbol{r} \tag{16}$$

where we define the partial derivative as

$$\frac{\partial^2 \rho(\boldsymbol{r})}{\partial E_j \partial Q_k} \equiv \chi_j(\boldsymbol{r})\,. \tag{17}$$

We evaluate this second derivative using the finite element method (FEM)

$$\chi_j(\boldsymbol{r}) = \frac{\left(\rho_{Q_+}^{E_j^+} - \rho_{Q_+}^{E_j^-}\right) - (\rho_{Q_-}^{E_j^+} - \rho_{Q_-}^{E_j^-})}{4\Delta Q_k \Delta E_j} \tag{18}$$

from the displaced geometries $Q_\pm = Q_0 \pm \Delta Q_k \boldsymbol{\varphi}_k$. $\boldsymbol{\varphi}_k$ is the real normal mode displacement vector which is not equal to the normalized normal mode displacement vector $\boldsymbol{\phi}_k$ a Gaussian16 calculation prints out. Joseph W. Ochterski[20] shows in his paper on Gaussian's vibrational analysis that the norm of each normal mode displacement vector is set to 1

$$\left|\boldsymbol{\phi}_{(k)}\right| = 1 \tag{19}$$

and the normal mode reduced masses are obtained from

$$\mu_k = \left(\sum_n^{\text{atoms}} \sum_{l=x,y,z} (\boldsymbol{\phi}_n^l)^2\right)^{-1}\,. \tag{20}$$

This normalization allows for an easy visualization of normal modes; however, the atomic displacements in different normal modes cannot be compared as their real displacements are scaled. Therefore, we remove the normalization factor to define the real normal mode displacement vector in cartesian coordinates.

$$\boldsymbol{\varphi}_n^l = \frac{\boldsymbol{\phi}_n^l}{\sqrt{\mu_k}} \tag{21}$$

Real displaced geometries can thus be calculated which is necessary to compare equivalent normal modes between molecules or different normal modes within one molecule.

All electron densities are calculated on a finite grid in a chosen cube. We find that the size and position of the cube as well as the choice of origin are crucial for stable FEM differentiation as especially the $\chi_j(\boldsymbol{r})$ maps are strongly origin dependent. A lot of work has been done in redefining origin independent molecular electronic properties[21,22], it would, however, go beyond the scope of this work to additionally establish origin independent RPD and RID maps. The $\chi_j'(\boldsymbol{r})$ normalization yields at least origin independent volume integrals of the RPDs. Assume a translation of the origin by the vector $\boldsymbol{a}$. This displaces all vectors by $\boldsymbol{r}' = \boldsymbol{r} - \boldsymbol{a}$. The electron densities remain at their original positions, $\chi_j'(\boldsymbol{r})$ and $d\boldsymbol{r}' = d\boldsymbol{r}$.

$$\int r_i' \, \chi_j'(\boldsymbol{r}) \, d\boldsymbol{r}' = \int (r_i - a_i) \, \chi_j'(\boldsymbol{r}) \, d\boldsymbol{r}' = \int r_i \, \chi_j'(\boldsymbol{r}) \, d\boldsymbol{r} - a_i \int \chi_j' \, (\boldsymbol{r}) \, d\boldsymbol{r} = \int r_i \, \chi_j'(\boldsymbol{r}) \, d\boldsymbol{r} \qquad (22)$$

Stable FEM differentiation is achieved from small perturbation so that we chose perturbation magnitudes of $\Delta Q_k = 0.05$ and $\Delta E_j = 0.001$ a.u. FEM differentiation was performed on a 80 x 80 x 80 points cube grid with step sizes of 0.5 Å. It spans from (-20,-20,-20) [Å] to (20,20,20) [Å]. The cube's origin was chosen as the Ag atom the N atom of MBT binds to in this SERS model, representing the center of the dominant molecule-substrate coupling region. This choice provides a physically motivated reference frame for interpreting the local electron density response in terms of the experimentally relevant adsorption geometry.

The obtained $R_{ij}(\boldsymbol{r})$ maps integrate to the respective Raman tensor entries

$$R_{ij} = -\int R_{ij}(\boldsymbol{r}) \, d\boldsymbol{r} \qquad (23)$$

except for a small numerical noise. At this point, we will neglect the factor $-1$ of the integral for clarity (consequently, large positive RPDs integrate to positive Raman tensor entries, which is more intuitive). The Raman Activity Density, RAD, is defined in the main text from the isotropic and anisotropic RPDs.

$$S_k(\boldsymbol{r}) = 45 R_{iso}(\boldsymbol{r}) + 7 R_{aniso}(\boldsymbol{r}) \qquad (24)$$

With eq. (23), we can show that the RAD integrates to the global Raman activity. For that, we consider the isotropic and anisotropic RPDs separately. The isotropic RPD integrates to the square of the reduced trace of the global Raman tensor.

$$\begin{aligned} \int R_{iso}(\boldsymbol{r}) \, d\boldsymbol{r} &= \frac{1}{3} \bar{\alpha}_k \left[ \int R_{xx}(\boldsymbol{r}) \, d\boldsymbol{r} + \int R_{yy}(\boldsymbol{r}) \, d\boldsymbol{r} + \int R_y(\boldsymbol{r}) \, d\boldsymbol{r} \right] \\ &= \frac{1}{3} \bar{\alpha}_k \left[ R_{xx} + R_{yy} + R_{zz} \right] \\ &= \bar{\alpha}_k^2 \end{aligned} \qquad (25)$$

Analogously, the anisotropic RPD integrates to the global anisotropy parameter (that is squared per definition).

$$\begin{aligned} \int R_{aniso}(\boldsymbol{r}) \, d\boldsymbol{r} &= \frac{1}{2} \int \Big[ \big( R_{xx}(\boldsymbol{r}) - R_{yy}(\boldsymbol{r}) \big) \left( R_{xx} - R_{yy} \right) + \big( R_{yy}(\boldsymbol{r}) - R_{zz}(\boldsymbol{r}) \big) \left( R_{yy} - R_{zz} \right) \\ &\qquad + \left( R_{zz}(\boldsymbol{r}) - R_{xx}(\boldsymbol{r}) \right) \left( R_{zz} - R_{xx} \right) \Big] \, d\boldsymbol{r} \end{aligned} \qquad (26)$$

$$+3\int\left[R_{xy}(\boldsymbol{r})R_{xy}+R_{xz}(\boldsymbol{r})R_{xz}+R_{yz}(\boldsymbol{r})R_{yz}\right]d\boldsymbol{r}$$

$$=\frac{1}{2}\Bigg[\left(\int R_{xx}(\boldsymbol{r})\,d\boldsymbol{r}-\int R_{yy}(\boldsymbol{r})\,d\boldsymbol{r}\right)\left(R_{xx}-R_{yy}\right)$$

$$+\left(\int R_{yy}(\boldsymbol{r})\,d\boldsymbol{r}-\int R_{zz}(\boldsymbol{r})\,d\boldsymbol{r}\right)\left(R_{yy}-R_{zz}\right)$$

$$+\left(\int R_{zz}(\boldsymbol{r})\,d\boldsymbol{r}-\int R_{xx}(\boldsymbol{r})\,d\boldsymbol{r}\right)\left(R_{zz}-R_{xx}\right)\Bigg]$$

$$+3\left[R_{xy}\int R_{xy}(\boldsymbol{r})\,d\boldsymbol{r}+R_{xz}\int R_{xz}(\boldsymbol{r})\,d\boldsymbol{r}+R_{yz}\int R_{yz}(\boldsymbol{r})\,d\boldsymbol{r}\right]$$

$$=\frac{1}{2}\left[\left(R_{xx}-R_{yy}\right)^2+\left(R_{yy}-R_{zz}\right)^2+\left(R_{zz}-R_{xx}\right)^2\right]+3\left[R_{xy}^2+R_{xz}^2+R_{yz}^2\right]$$

$$=\bar{\gamma}_k^2$$

Consequently, the RAD integrates to the global Raman activity

$$\begin{aligned}\int S_k(\boldsymbol{r})\,d\boldsymbol{r}&=45\int R_{iso}(\boldsymbol{r})d\boldsymbol{r}+7\int R_{aniso}(\boldsymbol{r})\,d\boldsymbol{r}\\&=45\bar{\alpha}_k^2+7\bar{\gamma}_k^2\\&=S_k\end{aligned}\tag{27}$$

and the Raman Intensity Density accordingly to the global Raman intensity.

Atomic RPDs can be analogously obtained from considering atomic displacement vectors. In matrix representation, the global displacement vector $\boldsymbol{\varphi}_k$ can be represented as

$$\boldsymbol{\varphi}_k=\begin{pmatrix}\varphi_{k,x}^{n_1}&\varphi_{k,y}^{n_1}&\varphi_{k,z}^{n_1}\\\varphi_{k,x}^{n_2}&\varphi_{k,y}^{n_2}&\varphi_{k,z}^{n_2}\\\vdots&\vdots&\vdots\\\varphi_{k,x}^{n_N}&\varphi_{k,y}^{n_N}&\varphi_{k,z}^{n_N}\end{pmatrix}\tag{28}$$

with the rows showing the x, y and z displacements of atom $n_i$. Consequently, atomic displacement tensors can be defined by extracting the row of the considered atom.

$$\boldsymbol{\varphi}_k^{n_1}=\begin{pmatrix}\varphi_{k,x}^{n_1}&\varphi_{k,y}^{n_1}&\varphi_{k,z}^{n_1}\\0&0&0\\\vdots&\vdots&\vdots\\0&0&0\end{pmatrix},\boldsymbol{\varphi}_k^{n_2}=\begin{pmatrix}0&0&0\\\varphi_{k,x}^{n_2}&\varphi_{k,y}^{n_2}&\varphi_{k,z}^{n_2}\\\vdots&\vdots&\vdots\\0&0&0\end{pmatrix},\cdots\tag{29}$$

With this definition, the atomic displacement vectors sum up to the global displacement vector.

$$\sum_n^N\boldsymbol{\varphi}_k^n=\boldsymbol{\varphi}_k\tag{30}$$

Atomic displaced geometries can thus be analogously obtained from these atomic displacement vectors. From these atomic displaced geometries, atomic RPDs can be analogously defined

$$R_{ij}^n(\boldsymbol{r})=r_i\chi_j^n(\boldsymbol{r})\tag{31}$$

with $\chi_j^n(\boldsymbol{r})$ being obtained from FEM differentiation after normalization corresponding to eq. (12) in the main text. These atomic RPDs are defined such that their integrals recover the corresponding atomic Raman tensor entries.

$$\int R_{ij}^{n}(\boldsymbol{r})d\boldsymbol{r} = R_{ij}^{n} \tag{32}$$

Atomic reduced trace RPDs are defined and used to define atomic isotropic RPDs

$$R_{iso}^{n}(\boldsymbol{r}) = \bar{R}^{n}(\boldsymbol{r})\bar{\alpha}_k \tag{33}$$

That integrate to the corresponding atomic isotropic Raman intensity contributions.

$$\begin{aligned}\int R_{iso}^{n}(\boldsymbol{r})\,d\boldsymbol{r} &= \bar{\alpha}_k \frac{1}{3}\left(\int R_{xx}^{n}(\boldsymbol{r})d\boldsymbol{r} + \int R_{yy}^{n}(\boldsymbol{r})d\boldsymbol{r} + \int R_{zz}^{n}(\boldsymbol{r})d\boldsymbol{r}\right)\\ &= \bar{\alpha}_k \frac{1}{3}\left(R_{xx}^{n} + R_{yy}^{n} + R_{zz}^{n}\right)\\ &= \bar{\alpha}_k^{n}\bar{\alpha}_k = a_k^{n}\end{aligned} \tag{34}$$

Atomic anisotropic RPDs can also be analogously defined

$$\begin{aligned}R_{aniso}^{n}(\boldsymbol{r}) &= \frac{1}{2}[\left(R_{xx}^{n}(\boldsymbol{r}) - R_{yy}^{n}(\boldsymbol{r})\right)\left(R_{xx} - R_{yy}\right) + \left(R_{yy}^{n}(\boldsymbol{r}) - R_{zz}^{n}(\boldsymbol{r})\right)\left(R_{yy} - R_{zz}\right)\\ &\quad + \left(R_{zz}^{n}(\boldsymbol{r}) - R_{xx}^{n}(\boldsymbol{r})\right)(R_{zz} - R_{xx})] + 3\left[R_{xy}^{n}(\boldsymbol{r})R_{xy} + R_{xz}^{n}(\boldsymbol{r})R_{xz} + R_{yz}^{n}(\boldsymbol{r})R_{yz}\right]\end{aligned} \tag{35}$$

that integrate to the corresponding atomic anisotropic Raman intensity contributions.

$$\begin{aligned}\int R_{aniso}^{n}(\boldsymbol{r})d\boldsymbol{r} &= \frac{1}{2}\Big[\left(\int R_{xx}^{n}(\boldsymbol{r})d\boldsymbol{r} - \int R_{yy}^{n}(\boldsymbol{r})d\boldsymbol{r}\right)\left(R_{xx} - R_{yy}\right)\\ &\quad + \left(\int R_{yy}^{n}(\boldsymbol{r})d\boldsymbol{r} - \int R_{zz}^{n}(\boldsymbol{r})d\boldsymbol{r}\right)\left(R_{yy} - R_{zz}\right)\\ &\quad + \left(\int R_{zz}^{n}(\boldsymbol{r})d\boldsymbol{r} - \int R_{xx}^{n}(\boldsymbol{r})d\boldsymbol{r}\right)\left(R_{zz} - R_{xx}\right)\Big]\\ &\quad +3\left[R_{xy}\int R_{xy}^{n}(\boldsymbol{r})d\boldsymbol{r} + R_{xz}\int R_{xz}^{n}(\boldsymbol{r})d\boldsymbol{r} + R_{yz}\int R_{yz}^{n}(\boldsymbol{r})d\boldsymbol{r}\right]\\ &= \frac{1}{2}\left[\left(R_{xx}^{n} - R_{yy}^{n}\right)\left(R_{xx} - R_{yy}\right) + \left(R_{yy}^{n} - R_{zz}^{n}\right)\left(R_{yy} - R_{zz}\right)\right.\\ &\quad \left. + \left(R_{zz}^{n} - R_{xx}^{n}\right)\left(R_{zz} - R_{xx}\right)\right] + 3\left[R_{xy}^{n}R_{xy} + R_{xz}^{n}R_{xz} + R_{yz}^{n}R_{yz}\right]\\ &= \gamma_k^{n}\end{aligned} \tag{36}$$

The atomic Raman Activity Density is determined from the atomic isotropic and anisotropic RPDs

$$S_k^{n}(\boldsymbol{r}) = 45R_{iso}^{n}(\boldsymbol{r}) + 7R_{aniso}^{n}(\boldsymbol{r}) \tag{37}$$

that integrates to the atomic Raman activity due to the integration behavior of the isotropic and anisotropic RPDs discussed above. Finally, the atomic Raman Intensity Density is obtained from the atomic RAD and the mode-dependent prefactor.

$$I_k^{n}(\boldsymbol{r}) = \frac{(2\pi)^4}{45}(\nu_L - \nu_k)^4 \cdot \frac{h}{8\pi^2 c\nu_k\left[1 - exp\left(-\frac{hc\nu_k}{kT}\right)\right]} \cdot S_k^{n}(\boldsymbol{r}) \tag{38}$$

It integrates accordingly to the atomic Raman intensity due to the atomic RAD integrating to the atomic Raman activity.

The computational effort is reduced in this work by defining the MBT molecular plane as xz-plane. Consequently, the Raman tensor entries in y-direction become negligible so that they can be neglected in the RID calculation. The (atomic) isotropic and anisotropic RPDs can be simplified as follows.

$$R_{iso}^{(n)}(\boldsymbol{r}) = \frac{1}{3}(R_{xx}^{(n)}(\boldsymbol{r}) + R_{zz}^{(n)}(\boldsymbol{r}))\bar{\alpha}_k \tag{39}$$

and

$$R_{aniso}^{(n)}(\boldsymbol{r}) \quad = \frac{1}{2}[R_{xx}^{(n)}(\boldsymbol{r})R_{xx} + R_{zz}^{(n)}(\boldsymbol{r})R_{zz} + \left(R_{zz}^{(n)}(\boldsymbol{r}) - R_{xx}^{(n)}(\boldsymbol{r})\right)(R_{zz} - R_{xx})] + 3R_{xz}^{(n)}(\boldsymbol{r})R_{xz}\,. \tag{40}$$

The (atomic) RADs and RIDs are calculated as shown in eqs. (37) and (38).

## S9. Normal mode RPDs

### S1. 1406 $cm^{-1}$ mode

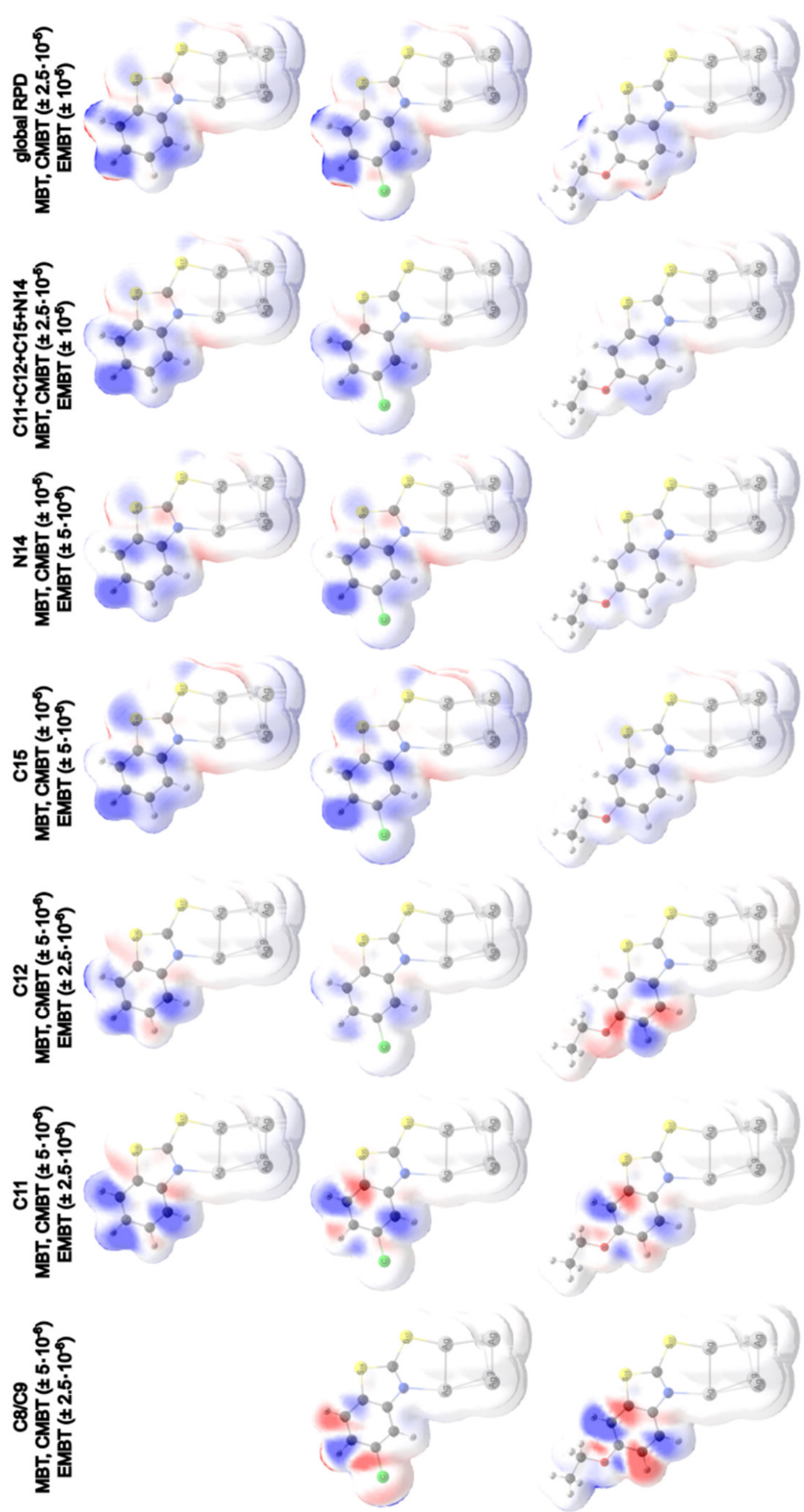

*Figure S7: (Atomic) RIDs of MBT, CMBT and EMBT for the 1406 $cm^{-1}$ peak. The color map min and max (in a.u. Å$^{-3}$) are given in parentheses.*

S2. 1463 cm$^{-1}$ mode

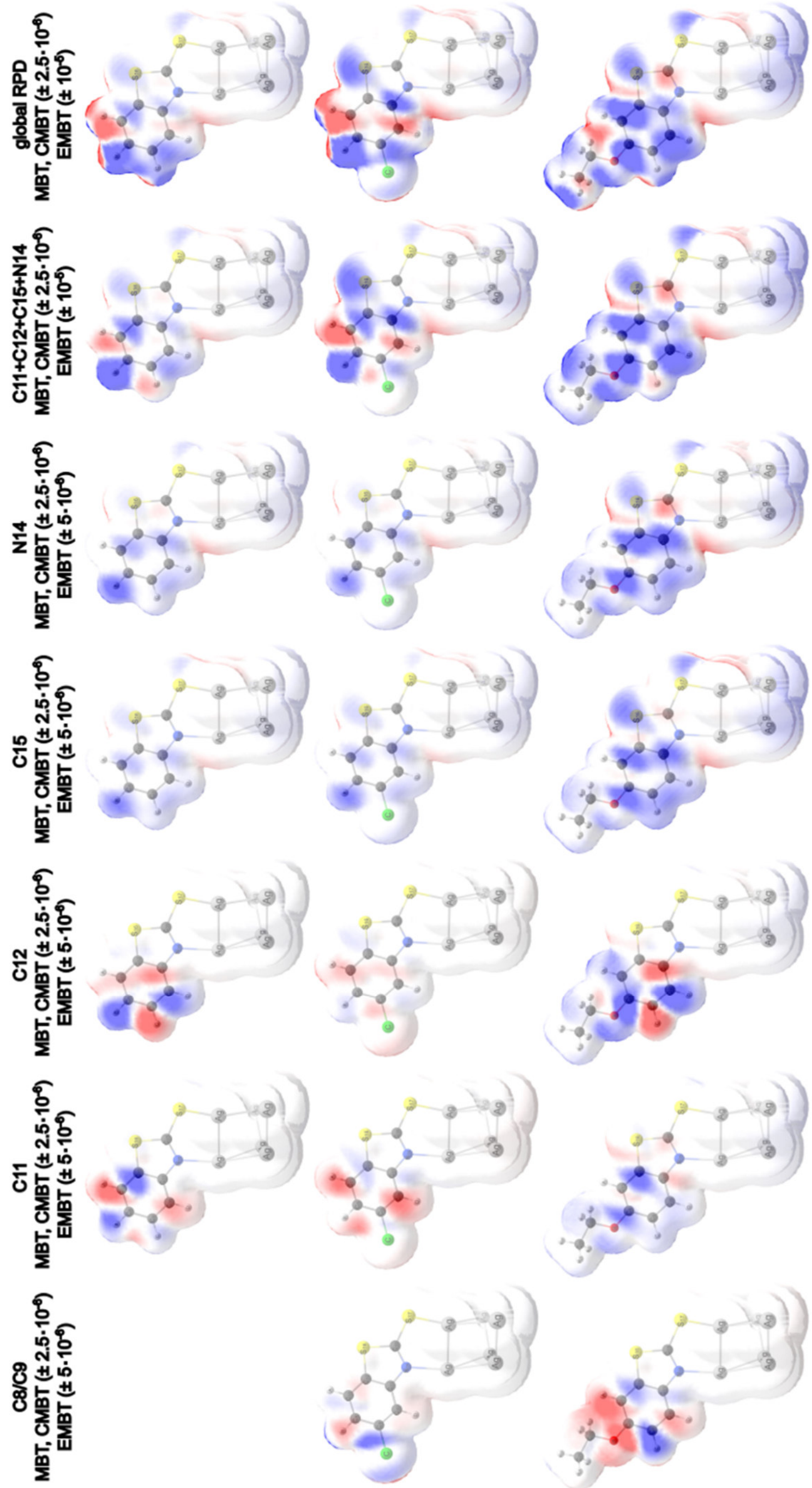

*Figure S8: (Atomic) RIDs of MBT, CMBT and EMBT for the 1463 cm$^{-1}$ peak. The color map min and max (in a.u. Å$^{-3}$) are given in parentheses.*

## S10. Normal mode CDDs

### S1. 1406 $cm^{-1}$ mode

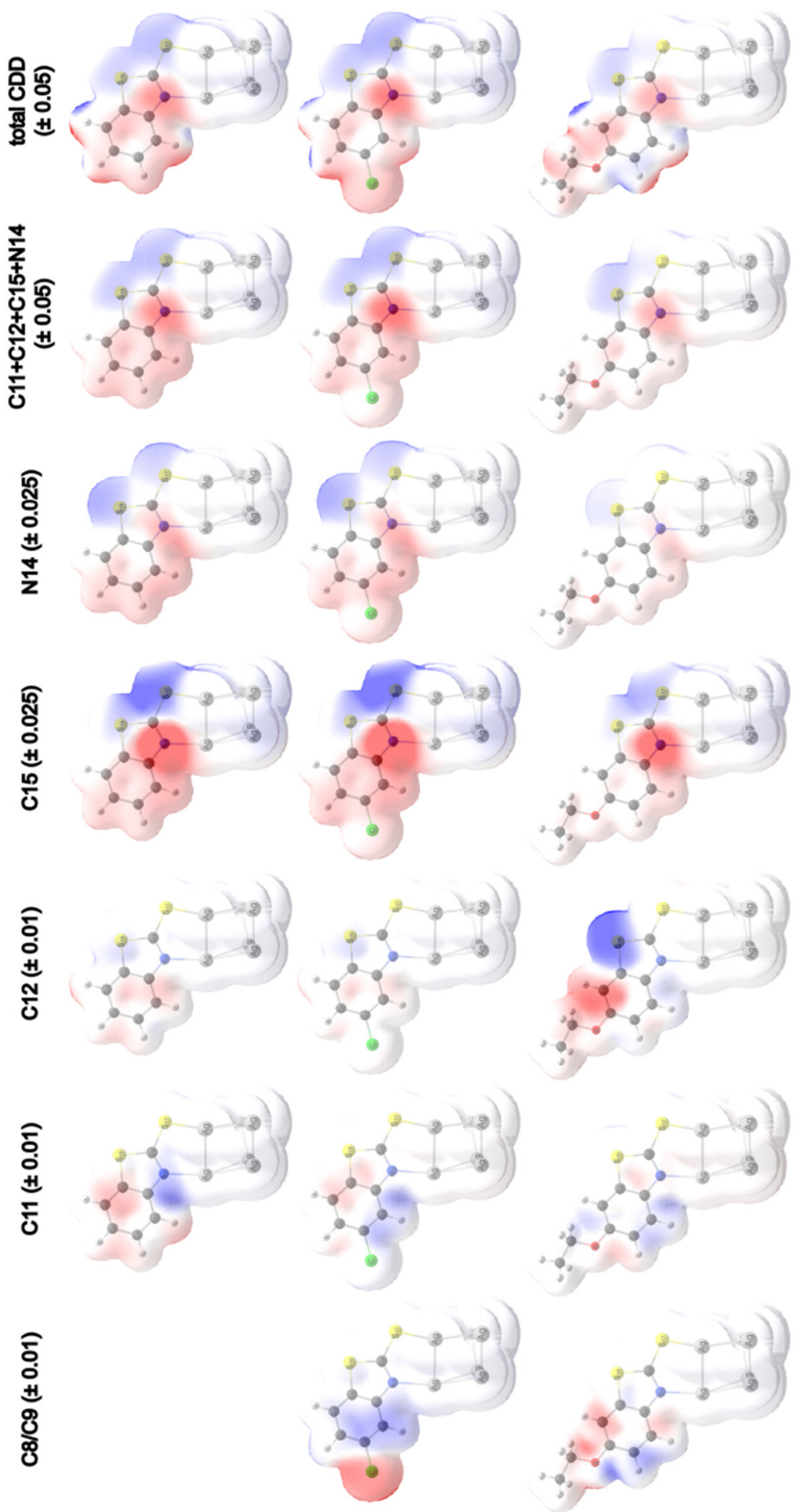

*Figure S9: (Atomic) CDDs of MBT, CMBT and EMBT for the 1406 $cm^{-1}$ peak. The color map min and max (in a.u.) are given in parentheses.*

S2. 1463 $cm^{-1}$ mode

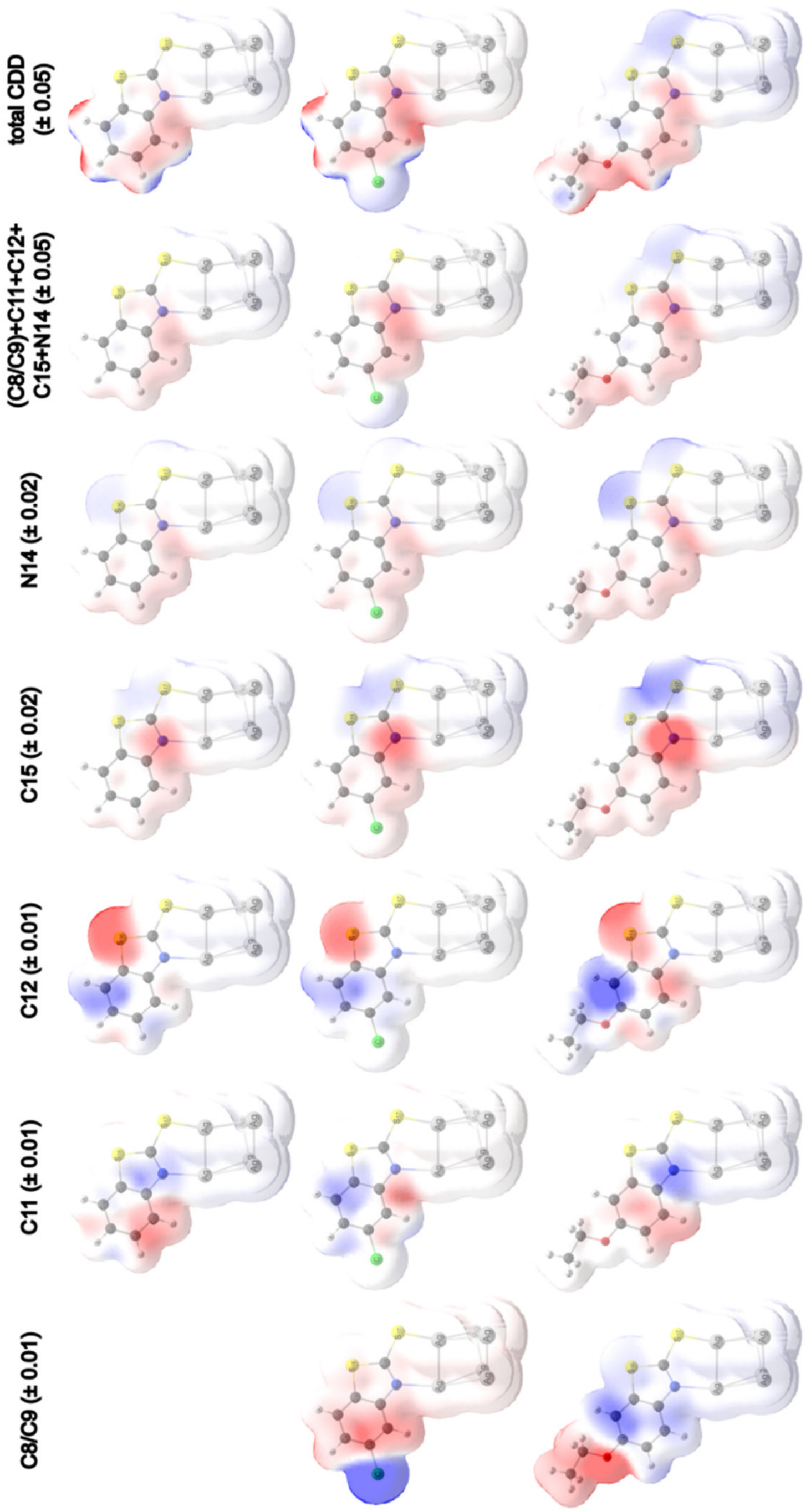

*Figure S10: (Atomic) CDDs of MBT, CMBT and EMBT for the 1406 $cm^{-1}$ peak. The color map min and max (in a.u.) are given in parentheses.*

## S11. Natural Population Analysis (NPA) of the 1406 $cm^{-1}$ and 1463 $cm^{-1}$ modes

*Table 1: NPA natural charges of the relevant MBT atoms.*

| | 1406 | | | 1463 | | |
|---|---|---|---|---|---|---|
| | MBT | CMBT | EMBT | MBT | CMBT | EMBT |
| C8 | -0.063 | -0.068 | -0.020 | -0.015 | 0.021 | -0.062 |
| C9 | -0.004 | 0.098 | -0.029 | 0.027 | -0.072 | 0.034 |
| C10 | -0.035 | -0.108 | -0.043 | 0.058 | 0.050 | 0.021 |
| C11 | 0.090 | 0.094 | 0.070 | -0.010 | -0.045 | 0.072 |
| C12 | -0.067 | -0.073 | -0.133 | 0.091 | 0.059 | 0.063 |
| C13 | 0.021 | 0.042 | 0.061 | -0.101 | -0.054 | -0.068 |
| N14 | -0.097 | -0.114 | -0.062 | -0.050 | -0.065 | -0.080 |
| C15 | -0.424 | -0.428 | -0.301 | -0.169 | -0.197 | -0.364 |
| S16 | 0.249 | 0.245 | 0.248 | -0.008 | -0.005 | 0.117 |
| S17 | 0.309 | 0.327 | 0.181 | 0.130 | 0.151 | 0.258 |

## S12. Normal Mode Oscillation of NBO E(2) energies

Weinhold's NBO method transforms the molecular orbitals of a system into a basis of localized bond orbitals (so-called natural bond orbitals, NBOs) yielding a Lewis-like description of the electronic structure. Within these structures, second-order perturbation theory can be used to calculate the stabilization energies of donor-acceptor interactions, the so-called E(2) energy. It is obtained via

$$E_{i\to j}^{(2)} = q_i \frac{F(i,j)^2}{\varepsilon_j - \varepsilon_i} \tag{41}$$

with $q_i$, the occupation of NBO $i$, the NBO Fock matrix element $F(i,j)$ describing the electronic interaction of NBOs $i$ and $j$ and the corresponding NBO orbital energies $\varepsilon_{i,j}$. Consequently, it is a measure of orbital interaction; the higher the stabilization energy E(2) of a NBO pair, the stronger their interaction, the larger the donation of occupancy from $i$ to $j$, the larger the deviation from the classical Lewis structure.

In this work, we calculate the E(2) energies for the most significant mesomeric conjugations into the benzene's $\pi$-system, p(N14) $\rightarrow$ $\pi^*$, p(S16) $\rightarrow$ $\pi^*$ and p(X) $\rightarrow$ $\pi^*$, with p(n) being the lone pair of atom n that is localized in the atom's p atom which is in conjugation with the benzene's $\pi$-system and X being Cl in CMBT and O in EMBT) for both extremal structures. Their differences, $\Delta$E(2), quantify the modulations of the corresponding donor-acceptor interactions describing how the respective mesomeric interaction is modulated along the considered normal mode. Table 2 shows the obtained $\Delta$E(2) values for the considered mesomeric interactions for the two normal modes of the three molecules.

*Table 2: $\Delta E(2)$ values for the considered mesomeric interactions obtained for the global 1406 $cm^{-1}$ and 1463 $cm^{-1}$ normal modes in the considered three molecules.*

| | 1406 $cm^{-1}$ | | | 1463 $cm^{-1}$ | | |
|---|---|---|---|---|---|---|
| | MBT | CMBT | EMBT | MBT | CMBT | EMBT |
| N14 → π* | 2.2 | 1.8 | -1.5 | -0.1 | 0.8 | 2.3 |
| S16 → π* | 0.3 | 0.2 | -0.1 | 0.3 | -1.6 | -0.9 |
| X → π* | | -0.4 | 0.6 | | 0.7 | -1.3 |

It shows that during the 1406 $cm^{-1}$ mode, the interaction of the lone pair of N14 and the benzene's π-system is significantly increased in MBT during the C15-N14 elongation consistent with the discussion in the main text: C15-N14 bond elongation induces a partial negative charge in N14 which is delocalized into the benzene's π-system via mesomeric conjugation. Analogously, the interaction of S16 and the benzene's π-system is also increased, but to a significantly lesser degree, which is likely due to the absence of the additional induced partial negative charge observed for N14. The modulation of the mesomeric effects of N14 (intrinsic mesomeric effect + induced partial negative charge) and S16 (exclusively intrinsic mesomeric effect) helps to explain the observed (atomic) CDDs, which are discussed in the main text. These effects do not change significantly with the Cl-substituent in CMBT. However, the Cl → π* interaction is also modulated in the normal mode. A decrease in the mesomeric conjugation of the Cl-substituent, combined with an increase in the mesomeric conjugation of N14 and S16, contributes to the observed electronic restructuring and its corresponding polarizability modulation, resulting in the discussed high RID. In EMBT, in contrast, both the N14 → π* and S16 → π* interactions are decreased (coinciding with the reported decreased CDD). The increased mesomeric conjugation of the EtO-substituent shows that, in this mode, the intrinsic mesomeric effect of the EtO-substituent counteracts the normal mode induced electronic restructuring yielding in sum a lower global CDD as well as RID.

For the 1463 $cm^{-1}$ mode, we observe only little modulation of the donor-acceptor interactions in MBT consistent with the low CDD and RID. The Cl-substituent changes this by increasing both the N14 → π* as well as the Cl → π* interaction while significantly decreasing the S16 → π* interaction. The simultaneous increase in N14 → π* and Cl → π* interactions may reduce the capacity of the π-system to accommodate an additional third electron-donating effect which could contribute to the observed decrease of the S16 → π* interaction. In EMBT, we see now that, in contrast to the 1406 $cm^{-1}$ mode, the O → π* interaction is significantly decreased while the N14 → π* interaction is significantly increased. The substituent's mesomeric effect now collaborates with the mode-induced electronic restructuring. This indicates that the reduced O → π* interaction allows the N14 induced electronic restructuring to act more effectively on the π-system, which, in total, leads to a significant change in the electronic structure during the oscillation. This results in a large change in polarizability, which is reflected in the large RID.

We additionally calculated the ΔE(2) modulations for the single atomic displaced geometries which are shown in Table 3. The E(2) energies are complex orbital interactions so that they cannot be decomposed into atomic contributions. The sum over the atomic E(2) values does consequently not yield the global E(2) value. However, the values are still valuable for the comparison of atomic effects in different molecules and normal modes.

*Table 3: ΔE(2) values for the considered mesomeric interactions obtained for the atomic 1406 cm-1 and 1463 cm-1 displaced geometries. Only the dominating mesomeric interactions are listed per atomic motion.*

| | | 1406 | | | 1463 | | |
|---|---|---|---|---|---|---|---|
| | | MBT | CMBT | EMBT | MBT | CMBT | EMBT |
| C15 | N14 → π* | 13.4 | 14.8 | 8.6 | 5.4 | 7.3 | 10.2 |
| N14 | N14 → π* | 24.1 | 23.0 | 12.0 | 12.1 | 12.0 | 21.0 |
| C11 | N14 → π* | 5.7 | -1.1 | -5.9 | 7.8 | -4.1 | 13.1 |
| C12 | S16 → π* | 5.5 | 5.5 | 18.4 | -14.9 | -17.7 | -7.5 |
| C8/C9 | X → π* | | -6.4 | 3.9 | | 9.5 | -15.7 |